\documentclass[AMA,STIX1COL]{WileyNJD-v2}

\articletype{Research Article}%

\received{6 April 2021}
\revised{20 July 2021}
\accepted{25 July 2021}

\usepackage[euler]{textgreek} % upright greek characters
\usepackage{hhline} % double hline without interupting vertical lines in tables
\usepackage{paralist} % allows in-paragraph enumerated lists
\usepackage{IEEEtrantools} % allows IEEEeqnarray for fine control of subnumbered equations
\usepackage{multirow} % allows more nuanced tables with split rows
\usepackage{setspace} % allows \setstretch in table environments

\definecolor{commentgray}{gray}{0.4}

\usepackage{enumitem}

\usepackage[binary-units=true]{siunitx}
\newcommand{\partialfrac}[2]{\frac{\partial #1}{\partial #2}}
\newcommand{\norm}[1]{\left\lVert#1\right\rVert}
\DeclareMathOperator*{\argmin}{arg\,min}
\DeclareMathOperator*{\mean}{mean}
\DeclareMathOperator*{\RMS}{RMS}
\DeclareMathOperator*{\SpecRad}{SpectralRadius}

\newcommand{\real}{\mathbb{R}}
\newcommand{\integer}{\mathbb{Z}}
\newcommand{\IJRNCCopyrightTextbox}[1]{% \placetextbox{<horizontal pos>}{<vertical pos>}{<stuff>}
  \setbox0=\hbox{#1}% Put <stuff> in a box
  \AddToShipoutPictureFG*{% Add <stuff> to current page foreground
    \put(\LenToUnit{0.075\paperwidth},\LenToUnit{0.59\paperheight}){\vtop{{\null}%
    \makebox{\begin{minipage}{2in}#1\end{minipage}}%
    }}%
  }%
}%

\newcommand{\eqlabel}[1]{\addtocounter{equation}{-1}\refstepcounter{equation}\label{#1}}
\floatname{algorithm}{Procedure}

\newcommand{\suspend}[1]{
\newcounter{#1}\setcounter{#1}{\value{enumi}}
}
\newcommand{\resume}[1]{
\setcounter{enumi}{\value{#1}}
}
\hypersetup{hidelinks} % remove border from cross reference links

\newcommand{\eLift}{\mathbf{e}}
\newcommand{\uLift}{\mathbf{u}}
\newcommand{\rLift}{\mathbf{r}}
\newcommand{\yLift}{\mathbf{y}}
\newcommand{\etaLift}{\textnormal{\textbf{\texteta}}}
\newcommand{\phiLift}{\textnormal{\textbf{\textphi}}}
\newcommand{\fLift}{\mathbf{f}}
\newcommand{\idx}{i}
\newcommand{\eigvalquant}{v}
\newcommand{\tdx}{\ell}
\newcommand{\jacobian}[2]{\partialfrac{#1}{#2}}
\newcommand{\ymod}{\hat{y}}
\newcommand{\yLiftmod}{\hat{\yLift}}
\newcommand{\ypreview}{\hat{\mathscr{y}}}
\newcommand{\ymeas}{y}
\newcommand{\xmod}{\hat{x}}
\newcommand{\flin}{\bar{f}}
\newcommand{\fmod}{\hat{f}}
\newcommand{\fmodsim}{\tilde{f}}
\newcommand{\fetasim}{\tilde{f}_{\eta}}
\newcommand{\hlin}{\bar{h}}
\newcommand{\hmod}{\hat{h}}
\newcommand{\gmod}{\hat{\mathbf{g}}}
\newcommand{\glin}{\bar{\mathbf{g}}}
\newcommand{\gtrue}{\mathbf{g}}
\newcommand{\etamod}{\hat{\eta}}
\newcommand{\etaLiftsim}{\tilde{\etaLift}}
\newcommand{\etamodsim}{\tilde{\eta}}
\newcommand{\Asim}{\tilde{A}}
\newcommand{\Bsim}{\tilde{B}}
\newcommand{\angtruth}{\psi}
\newcommand{\Mc}{M_c}
\newcommand{\Mp}{M_p}
\newcommand{\Mcmod}{\hat{M}_c}
\newcommand{\Mpmod}{\hat{M}_p}
\newcommand{\dctruth}{d_c}
\newcommand{\dptruth}{d_p}
\newcommand{\dcmod}{\hat{d}_c}
\newcommand{\dpmod}{\hat{d}_p}
\newcommand{\force}{c}
\newcommand{\kmod}{\hat{\kappa}}
\newcommand{\rPerceived}{r^*}
\newcommand{\unoise}{\omega_\force}
\newcommand{\ynoise}{\omega_y}
\newcommand{\grav}{\mathscr{g}}
\newcommand{\param}{\theta}
\newcommand{\parammod}{\hat{\param}}
\newcommand{\convset}{\mathcal{C}}
\newcommand{\convRate}{\mathcal{R}}
\newcommand{\mfinal}{m_{\text{final}}}

\begin{document}

\title{%
Iterative learning control with discrete-time nonlinear nonminimum phase models via stable inversion%
}

\author[1]{Isaac A. Spiegel*}

\author[2]{Nard Strijbosch}

\author[2]{Tom Oomen}

\author[1]{Kira Barton}

\authormark{SPIEGEL \textsc{et al}}

\address[1]{\orgdiv{Mechanical Engineering Department}, \orgname{University of Michigan Ann Arbor}, \orgaddress{\state{Michigan}, \country{USA}}}

\address[2]{\orgdiv{Mechanical Engineering Department}, \orgname{Eindhoven University of Technology}, \orgaddress{\state{North Brabant}, \country{The Netherlands}}}

\corres{*Isaac A. Spiegel, Mechanical Engineering Department, University of Michigan Ann Arbor. \email{ispiegel@umich.edu}}

\abstract[Summary]{

Output reference tracking can be improved by iteratively learning from past data to inform the design of feedforward control inputs for subsequent tracking attempts. This process is called iterative learning control (ILC).
This article develops a method to apply ILC to systems with nonlinear discrete-time dynamical models with unstable inverses (i.e. discrete-time nonlinear non-minimum phase models).
This class of systems includes
piezoactuators, electric power converters, and manipulators with flexible links,
which
may be found in nanopositioning stages, rolling mills, and robotic arms,
respectively.
As these devices may be required to execute
fine transient reference tracking tasks
repetitively
in contexts such as manufacturing, 
they may benefit from ILC.
Specifically, this article facilitates ILC of such systems by presenting
a new ILC synthesis framework that allows combination of the principles of
Newton's root finding algorithm
 with stable inversion, a technique for generating stable trajectories from unstable models.
The new framework, called Invert-Linearize ILC (ILILC),
is validated in simulation on a cart-and-pendulum system with model error, process noise, and measurement noise.
Where 
preexisting Newton-based ILC
diverges, 
ILILC  with stable inversion
converges, 
and does so in less than one third the number of trials necessary for the convergence of a gradient-descent-based ILC technique used as a benchmark.

}


\jnlcitation{\cname{%
\author{Spiegel IA}, 
\author{Strijbosch N}, 
\author{Oomen T},
and 
\author{Barton K}}.
\ctitle{Iterative learning control of discrete-time nonlinear nonminimum phase models via stable inversion}, \cjournal{Int J Robust Nonlinear Control}, \cyear{2021}.
\href{https://doi.org/10.1002/rnc.5726}{https://doi.org/10.1002/rnc.5726}}

\maketitle

\footnotetext{\textbf{Abbreviations:} ILC, iterative learning control; NILC, Newton iterative learning control; ILILC, invert-linearize iterative learning control}

% COPYRIGHT NOTIFICATION
\IJRNCCopyrightTextbox{
\footnotesize \textcopyright 2021 John Wiley \& Sons Ltd. Personal use of this material is permitted.
Permission from John Wiley \& Sons Ltd. must be obtained for all other uses, 
including 
creating new collective works, 
for resale or redistribution to servers or lists, 
or reuse of any copyrighted component of this work in other works.
This document is an accepted version. Full citation to published version:
\\
Spiegel IA, Strijbosch N, Oomen T, Barton K. Iterative learning control with discrete-time nonlinear nonminimum phase models via stable inversion. Int J Robust Nonlinear Control.
2021. \href{https://doi.org/10.1002/rnc.5726}{https://doi.org/10.1002/rnc.5726}
}

% Outline tex file that contains input commands for all body tex files
\section{Introduction}
Iterative learning control (ILC) is the process of learning an optimal feedforward control input over multiple trials of a repetitive process based on 
feedback measurements from previous trials. 
Compared to real-time-feedback and/or feedforward control techniques, many case studies of ILC have shown a substantial reduction in tracking error.
Relevant
applications
include robot-assisted stroke rehabilitation \cite{Freeman2015}, high speed train control \cite{Yu2018}, laser additive manufacturing \cite{Rafajlowicz2019}, 
and vehicle-mounted manipulators \cite{Xing2019},
all of which use nonlinear 
models.
In fact, while the majority of ILC literature focuses on linear systems, the prevalence of nonlinear dynamics in 
real-world systems
has motivated the development of numerous ILC 
theories for
discrete-time nonlinear models 
\cite{Jang1994,Saab1995,Wang1998,Sun2003,Zhang2019,Xing2021}.

\setcitestyle{citesep={, }} % for textual citenum
In addition to the state nonlinearities most commonly treated by nonlinear systems literature, many real-life systems exhibit dynamics well-represented by models with at least one of the following properties: 
\begin{inparaenum}[(P1)]
\item
\label{R1}
relative degree $\geq 1$,
\item
\label{R2}
input nonlinearities,
\item
\label{R3}
time-variation, and
\item
\label{R4}
instability of the model inverse.
\end{inparaenum}
For example,
(P\ref{R1}) may be 
exhibited
in the position control of myriad systems including piezoactuators \cite{Shieh2008}, motors \cite{Hackl2013}, robotic manipulators \cite{Geniele1997}, and vehicles \cite{Munz2011}.
(P\ref{R2}) may be 
exhibited
by piezoactuators \cite{Shieh2008}, electric power converters \cite{Escobar1999}, wind energy systems \cite{DeBattista2004}, magnetic levitation systems \cite{Gutierrez2005}, and flexible-link manipulators \cite{Geniele1997}.
(P\ref{R3}) may be 
exhibited
by any feedforward-input-to-output model of systems using both feedforward and feedback control, as is often done for robotic manipulation \cite{Khosla1988}.
Finally, and of primary concern in this work, (P\ref{R4}) may be 
exhibited
by piezoactuators \cite{Schitter2002}, electric power converters \cite{Escobar1999}, wind energy systems \cite{DeBattista2004}, DC motor and tachometer assemblies \cite{Awtar2004}, and flexible-link manipulators \cite{Geniele1997}.
However, published
discrete-time-nonlinear-model-based ILC 
theories
exclude at least one of properties (P\ref{R1})-(P\ref{R4}) from consideration.
While the prior art makes important contributions such as foundational nonlinear ILC theory\cite{Saab1995,Jang1994,Wang1998}, relaxation of process repetitiveness assumptions\cite{Sun2003}, robustness to packet dropout in measurement and controller signals\cite{Zhang2019}, and integration of ILC with adaptive control\cite{Xing2021}, these studies' analyses are limited to specific system structures.
As a consequence, (P\ref{R1}) is not addressed by references \citenum{Saab1995,Wang1998,Zhang2019}, (P\ref{R2}) is not addressed by references \citenum{Jang1994,Saab1995,Wang1998,Zhang2019,Xing2021}, (P\ref{R3}) is not addressed by references \citenum{Jang1994,Sun2003}, and (P\ref{R4}) is not addressed by references \citenum{Jang1994,Saab1995,Wang1998,Sun2003,Zhang2019}
\footnote{
References \citenum{Jang1994,Saab1995,Wang1998,Sun2003,Zhang2019} do not explicitly discuss inverse instability issues, but 
the appendix
shows that the ILC schemes they present may fail to converge for systems with unstable inverses.
}.
\setcitestyle{citesep={,}} % reset to default for superscript citations

The fact that many of 
the above
example systems 
exhibit multiple properties
and many of the above ILC theories exclude multiple properties from consideration
illustrates that it can be challenging 
to find a model-based ILC synthesis scheme appropriate for 
many real-world applications.
Indeed, flexible-link manipulators 
exhibit all four properties,
and they are relevant to the fast and cost-effective automation of pick-and-place and assembly tasks as well as to the control of large structures such as cranes \cite[ch. 6]{DeWit1996}. 
Such application spaces would
benefit from having
a versatile ILC scheme 
compatible with
(P\ref{R1})-(P\ref{R4}).

Additionally, while
ILC seeks to converge to a satisfactorily low error, this learning is not immediate, and trials executed before the satisfactory error threshold is passed may be seen as costly failures from the perspective of the process 
specification.
It is thus desirable to develop ILC schemes that converge as quickly as possible.

One ILC scheme that comes close to meeting these needs for versatility and speed is
that of Avrachenkov\cite{Avrachenkov1998}, called Newton ILC  (NILC) here.
NILC  is
the application of Newton's root finding algorithm to a complete finite time series (as opposed to individual points in time).
NILC's synthesis procedure and convergence analysis 
are unusually broad in that they 
admit discrete-time nonlinear models with properties
(P\ref{R1})-(P\ref{R3}) 
\cite{Avrachenkov1998,Spiegel2019}.
Additionally, 
Newton's method has been shown to deliver faster convergence in ILC than 
schemes such as P-type ILC \cite[ch. 5]{Xu2003},
upon which much of the relevant prior art on the ILC of discrete-time nonlinear systems is founded \cite{Jang1994,Saab1995,Wang1998,Sun2003,Zhang2019}.
However, this work demonstrates that when synthesized from models with unstable inverses, i.e. non-minimum  phase models, NILC  typically generates control signals that diverge towards 
very large
magnitudes. 
In other words NILC 
may be incompatible with models exhibiting (P\ref{R4}).
This article presents
a new ILC framework inheriting the benefits of 
NILC  while surmounting this shortcoming.

For linear 
 models with unstable inverses,  
a
common way to obtain feedforward control signals 
is
to systematically synthesize 
approximate dynamical  
models
with stable inverses 
by individually changing the model zeros and poles, e.g. the work of Tomizuka\cite{Tomizuka1987}.
However, it is difficult to prescribe analogous systematic approximation methods for nonlinear models
because the poles and zeros do not necessarily manifest as distinct binomial factors 
in the system transfer function
that can be individually inverted or modified.

An alternative 
is to 
harness the fact that a scalar difference equation that is unstable 
when evolved forward in time from an initial condition
is stable if evolved backwards in time from a terminal 
condition.
If the stable and unstable modes of a system are decoupled and evolved in opposite 
directions, a stable total trajectory can be obtained.
This process is called 
stable inversion.
For linear systems on a bi-infinite timeline, with boundary conditions at time $\pm\infty$, stable inversion gives an exact solution to the output tracking problem posed by the unstable inverse model.
In practice on a finite timeline, a high-fidelity approximation is obtained by ensuring the reference is designed with sufficient room for pre- and post-actuation, i.e. with a ``flat'' beginning and end.
However,
unlike ILC, stable inversion alone cannot account for model error.
To address this,
Zundert et al
\cite{Zundert2016} 
details
stable inversion and presents an ILC scheme for linear systems that 
incorporates
a 
process similar to stable inversion.

Extension of stable inversion to nonlinear models
involves additional complexities.
Some of these challenges, e.g. the difficulty of completely decoupling the stable and unstable parts of a nonlinear system, have been addressed by works such as those of Devasia et al\cite{Devasia1996, Devasia1998} 
for continuous-time systems and Zeng et al\cite{Zeng2000} for discrete-time systems.
However, 
the following
challenges remain.
First, this prior art assumes that if the state and input are both zero at a particular time step, then the state will be zero at the next time step.
This is not true for most representations of systems employing both feedback and feedforward control
because 
if the reference is nonzero it drives state change via the feedback controller despite the initial state and feedforward input being zero.
Stable inversion erroneously based on this assumption can have poor performance, and stable inversion has not been proven to converge when this assumption is relaxed.
Secondly,
Zeng et al\cite{Zeng2000} 
does not translate
from the theoretical solution on a bi-infinite timeline to an implementable solution on a finite timeline.
This work addresses these challenges.

In short, 
although
the work to date on NILC  and stable inversion has made great strides, 
gaps remain between the prior art and
a synthesis scheme for ILC that is fast and 
applicable to a wide variety of models---including nonlinear non-minimum  phase models.
This leads to the 
main
contribution of 
the present article:
an
ILC framework 
enabling controller synthesis from models satisfying all of (P\ref{R1})-(P\ref{R4}).
The key elements of this framework are
\begin{itemize}[leftmargin=*]
    \item
    reversing the order
    of the linearization and model inversion processes in 
    NILC  to circumvent issues associated with matrix inversion,
    \item
    reformulation of the model inversion in NILC  as stable inversion,
    \item
    proof of stable inversion convergence with relaxed assumptions on state dynamics, enabling treatment of a wider array of feedback control and other time-varying models,
    and
    \item
    development of a structured method for implementing the stable inversion technique proposed in this work.
\end{itemize}
The
proposed framework is validated in simulation on a nonlinear, relative degree 2, time-varying, non-minimum  phase cart-and-pendulum system with model error and process and measurement noise.

The remainder of the paper is organized as follows. 
Section \ref{sec:back} provides technical details from the prior art in NILC  \cite{Spiegel2019} and stable inversion \cite{Zeng2000} necessary to present the novel contributions of the present work. 
Section \ref{sec:NILC} 
presents analysis that justifies
the attribution of a class of NILC  failures to 
inverse instability, 
and provides a new 
ILC framework
that enables the circumvention of this failure mechanism by incorporating stable inversion.
Section \ref{sec:StabInv} provides proof of convergence of stable inversion for an expanded class of systems and provides improved methods for practical implementation.
Section \ref{sec:Val} details and discusses the validation of the new 
ILC
framework
with stable inversion through benchmark simulations on a non-minimum  phase cart-and-pendulum system.
This includes demonstration of 
conventional
NILC's divergence when applied to the same system.
Section \ref{sec:conc} presents conclusions and areas for future work.

\section{Background}
\label{sec:back}

\subsection{Newton ILC}
\label{sec:NILCprior}
Consider
SISO, discrete-time, nonlinear, time-varying models 
\begin{IEEEeqnarray}{RL}
\eqlabel{eq:modelall}
\IEEEyesnumber
\IEEEyessubnumber*
\xmod_{\tdx}(k+1) &= \fmod\left(\xmod_\tdx(k),u_\tdx(k),k\right)
    \qquad \xmod_\tdx(0)=x_0 \quad \forall \tdx
    \label{eq:modelstate}
    \\
    \ymod_\tdx(k) &= \hmod(\xmod_\tdx(k))
    \label{eq:modelout}
    \\
    \IEEEnosubnumber*
    k&\in\{0,1,\dots,N\}
    \label{eq:krange}
\end{IEEEeqnarray}
where $\xmod\in\real^{n_x}$ is the state vector, 
$u\in\real$ is the control input,
$\ymod\in\real$ is the output,
and
$k$ is the discrete time index.
The system is made to perform repeated trials of a reference tracking task, where
$N\in\integer_{>0}$ is the number of time steps in a trial (i.e. the number of samples minus 1),
and
$\tdx\in\integer_{\geq0}$ is the trial index\footnote{
$\ell$ is used for the trial index because $i$ and $j$ will be used for matrix element indexing, $k$ is used for the discrete time index, $t$ is avoided to prevent confusion with continuous time, and $\ell$ is the next letter in the alphabet and thus commonly used for indexing.
}.
Additionally, consider the trial-invariant reference $r(k)\in\real$.
Hats, $\hat{}\,$, are used to emphasize that (\ref{eq:modelall}) is an imperfect model of some true system.
It is assumed that the control input and trial-invariant initial condition are perfectly known.

A
classical ILC structure
is given by
\begin{equation}
    \uLift_{\tdx+1}=\uLift_\tdx+L_\tdx\eLift_\tdx
    \label{eq:ILCclassic}
\end{equation}
where $\uLift\in\real^{N-\mu+1}$ and $\eLift\in\real^{N-\mu+1}$ are input and error time series vectors, $\mu$ is the relative degree of (\ref{eq:modelall}), and $L\in\real^{N-\mu+1\times N-\mu+1}$ is the learning 
matrix,
which must be designed by a human or generated by an automatic synthesis procedure.
The
time series vectors, also called lifted vectors, 
are explicitly given by
\begin{align}
    \uLift_\tdx&=\begin{bmatrix}
    u_\tdx(0) &
    \cdots & u_\tdx(N-\mu)
    \end{bmatrix}^T
    \label{eq:uvec}
    \\
    \eLift_\tdx=\rLift-\yLift_\tdx&=\begin{bmatrix}
    r(\mu)-\ymeas_\tdx(\mu) &
    \cdots & r(N)-\ymeas_\tdx(N)
    \end{bmatrix}^T
\end{align}
where $y\in\real$ is the measured output of the true, but unknowable,
system. 
These
unknown system
dynamics
are
represented as
the function
$\gtrue:\real^{N-\mu+1}\rightarrow\real^{N-\mu+1}$, which takes in $\uLift_\tdx$ and outputs $\yLift_\tdx$.

The work of Avrachenkov\cite{Avrachenkov1998} analyzes the convergence of (\ref{eq:ILCclassic}) within a ball around the solution input $\uLift_\text{d}$ (``solution'' meaning that $\gtrue\left(\uLift_\text{d}\right)=\rLift$). 
In the present context 
this ball can be defined as $S\left(\uLift_\text{d},\rho\right)=\{ \uLift\in\real^{N-\mu+1} | \norm{\uLift - \uLift_\text{d}}_2 < \rho\}$ with $\rho > 0$ and $\norm{\cdot}_2$ being the Euclidean norm.
Three conditions are posited:
\begin{enumerate}[label=(C\arabic*)]
\item
\label{C1}
The true dynamics $\gtrue$ are continuously differentiable with respect to $\uLift$ in $S\left(\uLift_\text{d},\rho\right)$ and their Jacobian $\jacobian{\gtrue}{\uLift}$ is Lipschitz continuous with respect to $\uLift$ in $S\left(\uLift_\text{d},\rho\right)$.
\item
\label{C2}
The learning matrix always has a bounded norm: 
$\norm{L_\tdx}_2<\varepsilon_1\in\real_{>0} \,\,\forall\,\tdx$
\item
\label{C3}
The learning matrix is sufficiently similar to the inverse of the true lifted system Jacobian:
$\norm{I-L_\tdx\jacobian{\gtrue}{\uLift}(\uLift_\tdx)}_2<1 \,\,\forall\,\tdx$
\suspend{conditions01}
\end{enumerate}
Avrachenkov proves\cite{Avrachenkov1998} that if \ref{C1}-\ref{C3} are satisfied within $S\left(\uLift_\text{d},\rho\right)$, then there exists a ball $S(\uLift_\text{d},\varepsilon_2)$ with $\varepsilon_2>0$ such that if
the initial guess $\uLift_0$ is an element of $S(\uLift_\text{d},\varepsilon_2)$
then (\ref{eq:ILCclassic}) converges to $\eLift=0_{N-\mu+1}$ as $\tdx\rightarrow\infty$.

NILC  is the use of the Newton-Raphson root finding algorithm
to derive an automatic synthesis formula for the trial-varying learning matrix $L_\tdx$.
The learning matrix is derived from the lifted representation of 
(\ref{eq:modelall})-(\ref{eq:krange}),
$\yLiftmod_\tdx=\gmod(\uLift_\tdx)$,
which is defined as follows.
Elements of $\yLiftmod_\tdx$ output by $\gmod$ are given via
\begin{align}
\label{eq:ymodk}
    \ymod_\tdx(k) = \hmod\left(\fmod^{(k-1)}(\uLift_\tdx)\right) \qquad k\in\{\mu,\mu+1,\cdots,N\}
\end{align}
where the parenthetical superscript notation indicates function composition of the form
\begin{align}
    \fmod^{(k)}(\uLift_\tdx)&=\fmod(\xmod_\tdx(k),u_\tdx(k),k)
    \\&=
    \fmod\left(\fmod\left(\vphantom{\fmod}\cdots,u_\tdx(k-1),k-1\right),u_\tdx(k),k\right)
\end{align}
Because 
$\xmod_\tdx(0)=x_0$ is known in advance and the time argument is determined by the element index of the lifted representation, $\yLiftmod_\tdx$ is a function of only $\uLift_\tdx$. Note that because the first element of $\yLiftmod_\tdx$ is $\ymod_\tdx(\mu)$ it explicitly depends on $u_\tdx(0)$.

Using
Newton's method to find the root of
the error time series
\begin{equation}
\eLift=\rLift-\yLift=\rLift-\gtrue(\uLift)
\end{equation}
yields
\begin{equation}
    L_\tdx=\left(\jacobian{\gtrue}{\uLift}(\uLift_\tdx)\right)^{-1}
    \label{eq:gammapriorimpossible}
\end{equation}
where $\jacobian{\gtrue}{\uLift}$ is the Jacobian 
of $\gtrue$ with respect to $\uLift$  as a function of $\uLift$.
This learning matrix formula is impossible to evaluate because of its dependence on the unknown dynamics $\gtrue$. Thus, $\gmod$ is used as an approximation of $\gtrue$ to yield the implementable NILC  learning matrix formula
\begin{equation}
    L_\tdx=\left(\jacobian{\gmod}{\uLift}(\uLift_\tdx)\right)^{-1}
    \label{eq:gammaprior}
\end{equation}

When NILC  was originally developed,
large Jacobians such as $\jacobian{\gmod}{\uLift}$ 
were
prohibitively difficult to derive and store as functions of $\uLift_\tdx$, necessitating the definition of additional approximation techniques. However, with
automatic differentiation tools such as CasADi \cite{Andersson2018}, the barrier to Jacobian computation is vastly reduced, and can be done directly in many cases.

\subsection{Stable Inversion}
\label{sec:stabinvprior}

The first step of stable inversion is 
deriving the conventional inverse.
To synthesize a minimal inverse system representation, first assume (\ref{eq:modelall})  is in the
normal
form
\begin{IEEEeqnarray}{RLR}
\eqlabel{eq:simmuall}
\IEEEyesnumber
\IEEEyessubnumber*
\label{eq:simmuminus}
    \xmod^{i}(k+1) &= \xmod^{i+1}(k)
    &
    \qquad
    i<\mu
\\
\xmod^{i}(k+1) &= \fmod^{i}\left(\xmod(k),u(k),k\right)
    &
    \qquad
    i\geq\mu
    \label{eq:simmu}
    \\
    \label{eq:simoutput}
    \ymod(k)&=\xmod^1
    &
\end{IEEEeqnarray}
where $\xmod(0)=0$, and the superscripts $i$ indicate the vector element index, starting from 1.
Note the ILC trial index subscript $\tdx$ is omitted in this section,  
as stable inversion on its own does not involve incrementing $\tdx$. 
Equation (\ref{eq:simmuminus}) captures the time delay arising from the system relative degree, while equation (\ref{eq:simmu}) captures the remaining system 
dynamics.
One method of deriving 
this 
normal
form from a system not in normal form is given in
Eksteen et al\cite{Eksteen2016}.
Note that this coordinate transformation is performed in advance of any stable inversion or ILC analysis or synthesis. Thus the coordinate transform does not interfere with satisfaction of the identical initial condition assumption in (\ref{eq:modelstate}).

Given this normal form,
use (\ref{eq:simoutput}) to
replace the first $\mu$ state variables with output variables via
\begin{align}
    \xmod^{i}(k) &= \ymod(k+i-1) \qquad i\leq\mu
    \label{eq:sub1}
\end{align}
Similarly, replace the $\mu$\textsuperscript{th} state variable incremented by one time step 
(i.e. the left side of (\ref{eq:simmu}) for $i=\mu$)
with an output variable via
\begin{align}
    \xmod^\mu(k+1)&= \ymod(k+\mu)
    \label{eq:sub2}
\end{align}
These substitutions are made to facilitate the inversion of system
(\ref{eq:simmuall}),
as the inverse of a system with relative degree $\mu\geq 1$ is necessarily acausal with dependence on some subset of $\{\ymod(k),\,\ymod(k+1),\cdots,\ymod(k+\mu)\}$ at each time step $k$.
For
notational compactness, define the $\ymod$-preview vector $\ypreview(k)\equiv[\ymod(k),\cdots,\ymod(k+\mu)]^T$.
Then inverting (\ref{eq:simmu}) with $i=\mu$ yields the conventional inverse output function
\begin{equation}
    u(k) = \fmod^{\mu}{\vphantom{\fmod}}^{ -1 }\left(
    \begin{bmatrix} \xmod^{\mu+1}, & \cdots, & \xmod^{n_x}  \end{bmatrix}^T,
    \ypreview(k),
    k\right)
    \label{eq:invoutx}
\end{equation}
where $\fmod^{\mu}{\vphantom{\fmod}}^{ -1 }$ is the inverse of $\fmod^{\mu}$, i.e. (\ref{eq:simmu}, $i=\mu$) solved for $u(k)$.
This output equation is substituted into 
(\ref{eq:simmu}, $i>\mu$)
along with (\ref{eq:sub1})-(\ref{eq:sub2})
to yield the entire inverse system dynamics
\begin{IEEEeqnarray}{RL}
\eqlabel{eq:invall}
\IEEEyesnumber
\IEEEyessubnumber*
\etamod(k+1) &= \fmod_{\eta}\left(\etamod(k),\ypreview(k),k\right)
    \label{eq:invstate}
    \\
    u(k)&=\fmod^{\mu}{\vphantom{\fmod}}^{-1}\left(\etamod(k),\ypreview(k),k\right)
    \label{eq:invout}
\end{IEEEeqnarray}
where $\etamod\in\real^{n_{\eta}}$ ($n_{\eta}=n_x-\mu$) is the inverse state vector defined
\begin{equation}
    \etamod^i(k)\equiv\xmod^{\mu+i}(k)
\end{equation}
and $\fmod_\eta:\real^{n_{\eta}}\times\real^{\mu+1}\times\integer\rightarrow\real^{n_{\eta}}$ is the inverse state dynamics 
\begin{equation}
    \fmod_\eta^i(\etamod(k),\ypreview(k),k)\equiv \fmod^{i+\mu}(\xmod(k),u(k),k)
\end{equation}

Next, 
a similarity transform is to be applied to
this inverse system 
to decouple the stable and unstable modes of its linearization about the initial condition. Consider the Jacobian
\begin{equation}
A=\jacobian{\fmod_{\eta}}{\etamod}\left(\etamod=0,\ypreview=\ypreview^\dagger,k=0\right)
\end{equation}
where $\ypreview^\dagger$ is the solution to $\fmod_{\eta}(0,\ypreview^\dagger,0)=0$. 
Then let $V$ be the similarity transform matrix such that
\begin{equation}
\label{eq:transform}
    \Asim=V^{-1}AV=
    \begin{bmatrix}
    \Asim_s & 0
    \\
    0
    & \Asim_u
    \end{bmatrix}
\end{equation}
where $\Asim_s\in\real^{\eigvalquant\times \eigvalquant}$ has all eigenvalues inside the unit circle, and $\Asim_u\in\real^{n_{\eta}-\eigvalquant\times n_{\eta}-\eigvalquant}$ has all eigenvalues outside the unit circle. This can be satisfied by deriving the real block Jordan form of $A$. 
The corresponding inverse system state dynamics are
\begin{equation}
    \etamodsim(k+1)=\fetasim\left(\etamodsim(k),\ypreview(k),k\right)
    \equiv V^{-1}\fmod_{\eta}\left(V\etamodsim(k),\ypreview(k),k\right)
    \label{eq:fetasim}
\end{equation}
where the tilde on $\fetasim$ indicates application to $\etamodsim$ rather than $\etamod$.
Note that despite using a linearization-derived linear similarity transform, (\ref{eq:fetasim}) describes the same nonlinear time-varying dynamics as (\ref{eq:invstate}), but with the linear parts of the stable and unstable modes decoupled.

If (\ref{eq:modelall})  
has an unstable inverse, 
then (\ref{eq:fetasim}) is unstable and $\etamodsim(k)$ will be unbounded as $k$ 
increases.
However, given an infinite timeline in the positive and negative direction, the equation
\begin{equation}
\label{eq:implicitsoln}
    \etamodsim(k) = \sum_{i=-\infty}^{\infty}
    \phi(k-i)\left(\fetasim\left(\etamodsim(i-1),\ypreview(i-1),i-1\right)-\Asim\etamodsim(i-1)\right)
\end{equation}
where
\begin{align}
    \phi(k) = \begin{cases}
    \begin{bmatrix}
    \Asim^k_s & 0_{\eigvalquant\times n_\eta-\eigvalquant} \\
    0_{n_\eta-\eigvalquant\times \eigvalquant} & 0_{n_\eta-\eigvalquant\times n_\eta-\eigvalquant}
    \end{bmatrix}
    & k>0
    \\[10pt]
    \begin{bmatrix}
    I_{\eigvalquant\times \eigvalquant} & 0_{\eigvalquant\times n_\eta-\eigvalquant} \\
    0_{n_\eta-\eigvalquant\times \eigvalquant} & 0_{n_\eta-\eigvalquant\times n_\eta-\eigvalquant}
    \end{bmatrix}
    & k=0
    \\[10pt]
    \begin{bmatrix}
    0_{\eigvalquant\times \eigvalquant} & 0_{\eigvalquant\times n_\eta-\eigvalquant} \\
    0_{n_\eta-\eigvalquant\times \eigvalquant} & -\Asim_u^k
    \end{bmatrix}
    & k<0
    \end{cases}
\end{align}
is an exact, bounded solution to (\ref{eq:fetasim}) provided the right hand side of (\ref{eq:implicitsoln}) exists for all $k\in\integer$. However, (\ref{eq:implicitsoln}) is implicit, and thus cannot be directly evaluated. A fixed-point problem
solver---Zeng et al\cite{Zeng2000}
uses Picard 
iteration---must
be used to find $\etamodsim$, and sufficient conditions for the solver convergence and solution uniqueness must be determined.

The Picard iterative solver \cite[ch. 9]{Agarwal2000} for (\ref{eq:implicitsoln}) is 
\begin{equation}
    \etamodsim_{(m+1)}(k) =
    \sum_{i=-\infty}^{\infty}
    \phi(k-i)\left(\fetasim\left(\etamodsim_{(m)}(i-1),\ypreview(i-1),i-1\right) - \Asim\etamodsim_{(m)}(i-1) \right)
    \label{eq:picardinf}
\end{equation}
where the parenthetical subscript $(m)\in\integer_{\geq 0}$ is the Picard iteration index.

To prove that (\ref{eq:picardinf}) converges to a unique solution, \cite{Zeng2000} makes the assumptions
that
\begin{enumerate}[label=(Z\arabic*)]
\item
\label{(Z2)}
$\fmod(0,0,k)=0$ $\forall$ $k$, and
\item
\label{(Z3)}
$\etamodsim_{(0)}(k)=0$ $\forall$ $k$.
\end{enumerate}
Note that
the continuous time literature also makes these assumptions
\cite{Devasia1996,Devasia1998}.

The 
first assumption
is violated for many representations of systems incorporating both feedback and feedforward control. An example of such a system is given in Section \ref{sec:Val}, where $u$ is the feedforward control input and the feedback control is part of the time-varying dynamics of $\fmod$. This feedback control influences $\xmod$ regardless of whether or not $u(k)=0$. While there may often be a change of variables that enables satisfaction of \ref{(Z2)},
(\ref{eq:simmuall})
already imposes constraints on the states and outputs, and for many systems it is unlikely for there to exist a change of variables satisfying both assumptions.

Furthermore, while for systems satisfying \ref{(Z2)}, \ref{(Z3)} may 
be the zero-input state trajectory, this is 
untrue
for systems violating \ref{(Z2)}. For these systems, the zero state trajectory \ref{(Z3)} is essentially arbitrary, and may 
degrade the quality of low-$m$ Picard iterates if far from the solution trajectory.
This 
jeopardizes convergence because the computational complexity of the Picard iteration solution grows exponentially with the number of iterations. It is thus desirable to reach a satisfactory solution in as few iterations as possible, i.e. it is desirable to have high-quality low-$m$ iterates. 

Section \ref{sec:StabInv} addresses these limitations by proving a new set of sufficient conditions for the unique convergence of (\ref{eq:picardinf}) that relaxes \ref{(Z2)}, \ref{(Z3)}.

\section{
ILC 
Analysis and Development
}
\label{sec:NILC}

In order to develop a new ILC framework for non-minimum phase models, it is necessary to concretely identify the failure mechanism of Section \ref{sec:NILCprior}'s NILC. Such analysis is absent in the literature, and is thus provided in Section \ref{sec:failure}. Section \ref{sec:gammanew} then presents a new learning matrix formula overcoming this failing. 

\subsection{
Failure for 
Models with Unstable Inverses
}
\label{sec:failure}
The 
NILC
scheme (\ref{eq:ILCclassic}), (\ref{eq:gammaprior})  provides convergence of $\eLift_\tdx$ to 0 in theory.
However, this assumes perfect computation of the matrix inversion in (\ref{eq:gammaprior}).
In practice, the precision to which $\left(\jacobian{\gmod}{\uLift}(\uLift_\tdx)\right)^{-1}$ can be accurately computed is directly dependent on the condition number of $\jacobian{\gmod}{\uLift}(\uLift_\tdx)$.
If the condition number 
of a matrix
is large enough, the values computed for 
its inverse
may become arbitrary, and their order of magnitude may grow directly with the order of magnitude of the condition number \cite{Rump2009}\textsuperscript{,}\cite[ch. 3.2]{Belsley1980}. This ``blowing up'' of the matrix inverse can cause divergence of (\ref{eq:ILCclassic}), (\ref{eq:gammaprior}).

In studies unrelated to NILC, large
 $\jacobian{\gmod}{\uLift}$ condition numbers have been 
 observed for non-minimum  phase linear systems, both time-invariant 
\cite{Chu2013}\textsuperscript{,}
\cite[ch. 5.3-5.4]{Moore1993} and time-varying \cite{Norrlof2002}\textsuperscript{,} \cite[ch. 4.1.1]{Dijkstra2004}. 
The fact that the minimum singular value of $\jacobian{\gmod}{\uLift}(\uLift_\tdx)$ decreases with increases in the system frequency response function magnitude at the Nyquist frequency \cite{Lee2000}
may contribute
to this ill-conditioning.
For linear systems, this magnitude is directly dependent on the zero 
magnitudes, and thus on the inverse systems' 
stability.

If inverse instability 
degrades the conditioning
of $\jacobian{\gmod}{\uLift}$ for linear models, it is guaranteed to do so for nonlinear models.
This is
because the Jacobian evaluated at a particular input trajectory, $\jacobian{\gmod}{\uLift}(\uLift^*)$, is equal to the constant matrix $\jacobian{\glin}{\uLift}$ where $\glin$ is the lifted input-output model of the linearization of (\ref{eq:modelall})  about the trajectory $\uLift^*$.

To illustrate this equality, first consider that the elements of $\jacobian{\gmod}{\uLift}(\uLift^*)$ are given by (\ref{eq:ymodk}) and the chain rule as
\begin{align}
\label{eq:jacobianelements}
    \partialfrac{\ymod(k)}{u(j)}(\uLift^*) =
    \partialfrac{\hmod}{\xmod}\left( \fmod^{(k-1)}(\uLift^*) \right)
    \partialfrac{\fmod^{(k-1)}}{\uLift}(\uLift^*)
    \partialfrac{\uLift}{u(j)}
\end{align}
where
\begin{align}
    \partialfrac{\uLift}{u(j)} = \begin{bmatrix}
    0_{1\times j} & 1 & 0_{1\times N-\mu+1-j}
    \end{bmatrix}^T
\end{align}
and $\partialfrac{\hmod}{\xmod}$ is a row vector.

Then consider the linearization of (\ref{eq:modelall})  about 
$\uLift^*$:
\begin{IEEEeqnarray}{RL}
\eqlabel{eq:linall}
\IEEEyesnumber
\IEEEyessubnumber*
\label{eq:flin}
    \delta \xmod(k+1) &=\flin\left(\delta\xmod(k),\delta u(k),k\right)
    = 
    \jacobian{\fmod}{\xmod}\left(\xmod^*(k),u^*(k),k\right)\,\delta \xmod(k) +
    \jacobian{\fmod}{u}(\xmod^*(k),u^*(k),k)\,\delta u(k)
    \\
    \label{eq:hlin}
    \delta \ymod(k) &= \hlin(\delta\xmod(k)) 
    =
    \jacobian{\hmod}{\xmod}(\xmod^*(k))\,\delta\xmod(k)
\label{eq:sysDef_y}
\end{IEEEeqnarray}
where
$\xmod^*(k)=\fmod^{(k-1)}(\uLift^*)$ and the $\delta$ notation denotes $\delta\xmod(k)=\xmod(k)-\xmod^*(k)$ for $\xmod$ and similar for $u$.

Lifting (\ref{eq:linall})  in the same manner as (\ref{eq:modelall})  yields the output perturbation as a function of the input perturbation time series $\delta\uLift$ via
\begin{align}
\label{eq:deltaymod}
    \delta \ymod (k) = \jacobian{\hmod}{\xmod}\left(\fmod^{(k-1)}(\uLift^*)\right)\flin^{(k-1)}(\delta\uLift)
\end{align}
Because of (\ref{eq:linall})'s linearity, $\flin^{(k-1)}(\delta\uLift)$ can be explicitly expanded as
\begin{align}
\label{eq:linfunccomp}
    \flin^{(k-1)}(\delta\uLift)=\left(\prod_{
    \kappa
    =0}^{k-1}\partialfrac{\fmod^{(
    \kappa
    )}}{\fmod^{(
    \kappa
    -1)}}(\uLift^*)\right)\delta\xmod(0) + \partialfrac{\fmod^{(k-1)}}{\uLift}(\uLift^*)\delta\uLift
\end{align}
where $\prod$ is ordered with the factor of least 
$\kappa$
on the right and the factor of greatest 
$\kappa$
on the left.
The terminal condition of the recursive function composition is $\fmod^{(-1)}=\xmod(0)$.
From (\ref{eq:deltaymod}) and (\ref{eq:linfunccomp}) it is clear that the elements of $\jacobian{\glin}{\delta\uLift}$ are given by
\begin{align}
    \partialfrac{\delta\ymod(k)}{\delta u(j)} = \jacobian{\hmod}{\xmod}\left(\fmod^{(k-1)}(\uLift^*)\right)
    \jacobian{\fmod^{(k-1)}}{\uLift}(\uLift^*)
    \jacobian{\delta\uLift}{\delta u(j)}
    ,
\end{align}
which is equal to (\ref{eq:jacobianelements}) because $\jacobian{\delta\uLift}{\delta u(j)}=\partialfrac{\uLift}{u(j)}$ 
due to the identical structures (\ref{eq:uvec}) of $\uLift$ and $\delta \uLift$ with respect to $u$ and $\delta u$ time indexing.
Thus, if (\ref{eq:modelall})  is such that its linearization (\ref{eq:linall})  is 
unstable, 
$\jacobian{\gmod}{\uLift}(\uLift_\tdx)$
will suffer ill-conditioning and  
attempts to compute
the learning matrix
(\ref{eq:gammaprior})
may
yield a matrix with large arbitrary elements. Such a learning gain matrix may in turn cause $\uLift_{\tdx+1}$ to contain large arbitrary elements, causing the learning law to diverge.

Therefore, for the learning law (\ref{eq:ILCclassic}) to converge for a 
system with an unstable inverse
in practice, a learning matrix synthesis that does not require matrix inversion of $\jacobian{\gmod}{\uLift}(\uLift_\tdx)$ is desired.

\subsection{
Alternative Learning Matrix Synthesis
}
\label{sec:gammanew}
To circumvent issues associated with inverting 
$\jacobian{\gmod}{\uLift}(\uLift_\tdx)$
a new learning matrix definition seeking to satisfy the requirements \ref{C2}-\ref{C3} in the spirit of Newton's method, but without the matrix inversion requirement of 
(\ref{eq:gammaprior}),
is given by
\begin{equation}
    L_\tdx = 
    \jacobian{\gmod^{-1}}{\yLiftmod}(\yLift_\tdx)
    \label{eq:gammanew}
\end{equation}
where $\gmod^{-1}:\real^{N-\mu+1}\rightarrow\real^{N-\mu+1}$ is a lifted model of the inverse of (\ref{eq:modelall}).
This makes 
$\jacobian{\gmod^{-1}}{\yLiftmod}$
a function of the output of (\ref{eq:modelall}), 
namely $\yLiftmod_\tdx$. As stated in Section \ref{sec:NILCprior}, 
$\yLiftmod_\tdx$ 
is the output of a necessarily erroneous model, and thus
is merely a prediction of the accessible, measured output $\yLift_\tdx$. Hence $\yLift_\tdx$ is used as the input to 
$\jacobian{\gmod^{-1}}{\yLiftmod}$.
In short, this work proposes using the linearization of the inverse of (\ref{eq:modelall})  rather than the inverse of the linearization,
and thus the new framework (\ref{eq:ILCclassic}), (\ref{eq:gammanew})  will be referred to as ``Invert-Linearize ILC'' (ILILC).

The first step in deriving $\gmod^{-1}$, and thus in deriving (\ref{eq:gammanew}) is the inversion of the original model (\ref{eq:modelall}) .
A direct method of inverting (\ref{eq:modelall})  is to solve
\begin{equation}
    \ymod_\tdx(k+\mu) = \hmod(
    \fmod^{(k+\mu-1)}(\uLift_\tdx)
    )
\end{equation}
for $u_\tdx(k)$, and substitute the resulting function of 
$\left\{\ymod_\tdx(k),\,\ymod_\tdx(k+1),\cdots,\ymod_\tdx(k+\mu) \right\}$
into (\ref{eq:modelstate}).
However, if (\ref{eq:modelall})  
has an unstable inverse, 
this method of inversion will yield unbounded states $\xmod_\tdx(k)$ as $k$ increases. Thus, $\gmod^{-1}$ is derived via
the
stable inversion
procedure described in Section \ref{sec:StabInv}
rather than direct inversion.
Note, though, that (\ref{eq:gammanew}) also admits the use of other stable approximate inverse models
for $\gmod^{-1}$ 
should they be available.

\section{
Stable Inversion Development
}
\label{sec:StabInv}
This section proves a relaxed set of sufficient conditions for the convergence of Picard iteration to the unique solution to the stable inversion problem, i.e. the unique solution to (\ref{eq:implicitsoln}) from Section \ref{sec:stabinvprior}.
This enables stable inversion---and thus ILC---for a new class of system representations capturing simultaneous feedback and feedforward control.
Additionally, a new initial Picard iterate prescription is given to suit the broadened scope of stable inversion, and a procedure for practical implementation is described.
This procedure enables the derivation of $\gmod^{-1}$.
\subsection{
Fixed-Point Problem Solution
}
Several definitions are 
needed to prove the relaxed set of sufficient conditions for convergence of the fixed-point problem solver used for stable inversion.

\begin{definition}[Lifted 
Matrices and Third-Order Tensors]
Given the vector and matrix functions of time $a(k)\in\real^{n}$ 
and
$B(k)\in\real^{n\times n}$, the corresponding lifted matrix and third order tensor are given by upright bold notation: $\mathbf{a}\in\real^{n\times\mathcal{K}}$ and $\mathbf{B}\in\real^{n\times n\times\mathcal{K}}$. $\mathcal{K}$ is the time dimension, and may be $\infty$. Elements of the lifted objects are $\mathbf{a}^{i,k}\equiv a^i(k)$ and $\mathbf{B}^{i,j,k}\equiv B^{i,j}(k)$.
\label{def:lift}
\end{definition}

\begin{definition}[Matrix and Third-Order Tensor Norms]
$\norm{\cdot}_\infty$ refers to the ordinary $\infty$-norm when applied to vectors, and is the matrix norm induced by the vector norm when applied to matrices (i.e. the maximum absolute row sum). Additionally, the entry-wise $(\infty,1)$-norm is defined for the matrices and third-order tensors $\mathbf{a}$ and $\mathbf{B}$ from Definition \ref{def:lift} as
\begin{align}
\norm{\mathbf{a}}_{\infty,1}
\equiv \sum_{k\in\mathcal{K}}\norm{a(k)}_\infty
\qquad
\norm{\mathbf{B}}_{\infty,1}\equiv \sum_{k\in\mathcal{K}}\norm{B(k)}_{\infty}
\end{align}
\end{definition}

\begin{definition}[Local Approximate Linearity
\cite{Devasia1996,Zeng2000}
]
$\fetasim$ is locally approximately linear in $\etamodsim(k)$ and its $\etamodsim(k)=0$ dynamics,
 in a closed $s$-neighborhood around $(\etamodsim(k)=0,\fetasim(0,\ypreview(k),k)=0)$,
 with Lipschitz constants $K_1, K_2>0$ if $\exists s>0$ such that for any vectors
\begin{itemize}
    \item
    $a(k)$, $b(k)\in\real^{n_\eta}$ with $\norm{\cdot}_\infty\leq s$ $\forall k$, and
    \item
    $\mathscr{a}(k)$, $\mathscr{b}(k)\in\real^{\mu+1}$ such that $\norm{\fetasim(0,\mathscr{a}(k),k)}_\infty$, $\norm{\fetasim(0,\mathscr{b}(k),k)}_\infty\leq s$ $\forall k$
\end{itemize}
the following is true $\forall k$
\begin{multline}
    \left\lVert
    \left(
    \fetasim(a(k),\mathscr{a}(k),k)-Aa(k)
    \right) 
    -
    \left(
    \fetasim(b(k),\mathscr{b}(k),k)-Ab(k)
    \right)
    \right\rVert_\infty
    \\
    \leq
    K_1\norm{
    a(k)-b(k)
    }_\infty+
    K_2\norm{
    \fetasim(0,\mathscr{a}(k),k)-\fetasim(0,\mathscr{b}(k),k)}_\infty
    \label{eq:lal}
\end{multline}
\label{def:lal}
\end{definition}

With these definitions 
a new set of sufficient conditions for Picard iteration convergence 
 is established.
Proof of this theorem shares the approach of Zeng et al\cite{Zeng2000} in establishing the Cauchy nature of the Picard sequence. It is also influenced by the proofs of Picard iterate local approximate linearity for continuous-time systems in Devasia et al\cite{Devasia1996}.

\begin{theorem}
The Picard iteration (\ref{eq:picardinf}) converges to a unique solution to (\ref{eq:fetasim}) if the following sufficient conditions are met.
\begin{enumerate}[label=(C\arabic*)]
\resume{conditions01}
    \item 
    \label{(C5)}
    $\norm{\etaLiftsim_{(0)}}_{\infty,1}\leq s$
    \item
    \label{(C5b)}
    $\forall k$ $\exists\ypreview(k)=\ypreview^\dagger(k)$ such that $\fetasim(0,\ypreview^\dagger(k),k)=0$
    \item
    \label{(C6)}
    $\fetasim$ is locally approximately linear in the sense of (\ref{eq:lal})
    \item
    \label{(C7)}
    $K_1\norm{\phiLift}_{\infty,1}<1$
    \item
    \label{(C8)}
    $\frac{\norm{\phiLift}_{\infty,1}K_2
    \norm{
    \tilde{\fLift}_\eta(0,\ypreview)
    }_{\infty,1}
    }{1-\norm{\phiLift}_{\infty,1}K_1}\leq s$
    \suspend{conditions02}
\end{enumerate}
where $\phiLift$ and $
\norm{
    \tilde{\fLift}_\eta(0,\ypreview)
    }_{\infty,1} =
    \sum_{k=-\infty}^\infty \norm{\fetasim(0,\ypreview(k),k)}_\infty
$
 are defined by Definition \ref{def:lift}.
\label{thm:picard}
\end{theorem}
\begin{IEEEproof}
Proof 
that (\ref{eq:picardinf}) converges to a unique fixed point begins with 
an induction showing
that $\etamodsim_{(m)}(k)$ remains in the locally approximately linear neighborhood $\forall$ $k$, $m$. The base case of this induction is given by \ref{(C5)}. Then under the premise 
\begin{equation}
\norm{\etaLiftsim_{(m)}}_{\infty,1}\leq s
\label{eq:premise}
\end{equation}
the induction proceeds as follows.
Here, ellipses indicate the continuation of a line of mathematics.

By 
the Picard iterative solver 
(\ref{eq:picardinf}):
\begin{equation}
    \norm{\etaLiftsim_{(m+1)}}_{\infty,1}=\sum_{k=-\infty}^\infty
    \left\lVert
    \sum_{\idx=-\infty}^\infty
    \phi(k-\idx)
    \vphantom{\sum_{\idx=-\infty}^\infty}
    \left(
    \fetasim\left(\etamodsim_{(m)}(\idx-1),\ypreview(\idx-1),\idx-1\right)
    -A\etamodsim_{(m)}(\idx-1)
    \right)
    \right\rVert_\infty
    \cdots
\end{equation}
By the triangle inequality:
\begin{equation}
    \cdots \leq \sum_{k=-\infty}^\infty \sum_{\idx=-\infty}^\infty
    \left\lVert
    \phi(k-\idx)
    \left(
    \fetasim\left(\etamodsim_{(m)}(\idx-1),\ypreview(\idx-1),\idx-1\right)
    -A\etamodsim_{(m)}(\idx-1)
    \right)
    \right\rVert_{\infty}
    \cdots
\end{equation}
By the fact that for matrix norms induced by vector norms $\norm{Ba}\leq\norm{B}\norm{a}$ for matrix $B$ and vector $a$:
\begin{equation}
    \cdots \leq \sum_{k=-\infty}^\infty \sum_{\idx=-\infty}^\infty
    \norm{
    \phi(k-\idx)
    }_\infty
    \norm{
    \fetasim\left(\etamodsim_{(m)}(\idx-1),\ypreview(\idx-1),\idx-1\right)
    -A\etamodsim_{(m)}(\idx-1)
    }_{\infty}
    \cdots
\end{equation}
\begin{equation}
    \cdots = \sum_{\idx=-\infty}^\infty
    \left\lVert
    \fetasim\left(\etamodsim_{(m)}(\idx-1),\ypreview(\idx-1),\idx-1\right)
    -A\etamodsim_{(m)}(\idx-1)
    \right\rVert_{\infty}
    \sum_{k=-\infty}^\infty
    \norm{
    \phi(k-\idx)
    }_\infty
    \cdots
\end{equation}
By the fact that $\sum_{k=-\infty}^\infty\norm{\phi(k-\idx)}_\infty$ has the same value $\forall \idx$
\begin{equation}
    \cdots = \norm{\phiLift}_{\infty,1}
    \sum_{\idx=-\infty}^\infty
    \norm{
    \fetasim\left(\etamodsim_{(m)}(\idx-1),\ypreview(\idx-1),\idx-1\right)
    -A\etamodsim_{(m)}(\idx-1)
    }_\infty
    \cdots
\end{equation}
By \ref{(C5b)}:
\begin{equation}
    \cdots = \norm{\phiLift}_{\infty,1}
    \sum_{\idx=-\infty}^\infty 
    \left\lVert
    \left(
    \fetasim\left(\etamodsim_{(m)}(\idx-1),\ypreview(\idx-1),\idx-1\right)
    -A\etamodsim_{(m)}(\idx-1)
    \right)
    -
    \left(
    \fetasim\left(0,\ypreview^\dagger(\idx-1),\idx-1\right)
    -A(0)
    \right)
    \right\rVert_\infty
    \cdots
\end{equation}
By \ref{(C6)}:
\begin{equation}
    \cdots \leq
    \norm{\phiLift}_{\infty,1} \sum_{\idx=-\infty}^\infty
    K_1\norm{\etamodsim_{(m)}(\idx-1)}_\infty +
    K_2\norm{\fetasim\left(0,\ypreview(\idx-1),\idx-1\right)}_\infty
    \cdots
\end{equation}
\begin{equation}
    \cdots =
    \norm{\phiLift}_{\infty,1}\left(
    K_1\norm{\etaLiftsim_{(m)}}_{\infty,1}
    +
    K_2\norm{\tilde{\fLift}_\eta(0,\ypreview)}_{\infty,1}
    \right)
    \cdots
\end{equation}
By \ref{(C7)}, 
the denominator of \ref{(C8)} is positive. Thus
both sides of \ref{(C8)} can be multiplied by 
this denominator
without changing the inequality direction. Thus by (\ref{eq:premise}) and algebraic rearranging of \ref{(C8)}
\begin{equation}
    \cdots \leq \norm{\phiLift}_{\infty,1}\left(
    K_1s+K_2\norm{\tilde{\fLift}_\eta(0,\ypreview)}_{\infty,1}
    \right)
    \leq
    s
\end{equation}
$\therefore$ $\norm{\etaLiftsim_{(m)}}_{\infty,1}\leq s$ $\forall m$. Because $\norm{\etaLiftsim_{(m)}}_{\infty,1}\geq \norm{\etamodsim_{(m)}(k)}_\infty$ $\forall k$, this implies that $\etamodsim_{(m)}(k)$ is within the locally approximately linear neighborhood $\forall$ $m$, $k$.

To show that (\ref{eq:picardinf}) converges to a unique fixed point, define
\begin{equation}
    \Delta\etamodsim_{(m)}(k)\equiv \etamodsim_{(m+1)}(k) - \etamodsim_{(m)}(k)
\end{equation}
Then, by a nearly identical induction
\begin{equation}
    \norm{\Delta\etaLiftsim_{(m+1)}}_{\infty,1} \leq \norm{\phiLift}_{\infty,1}K_1\norm{\Delta\etaLiftsim_{(m)}}_{\infty,1}
\end{equation}
By \ref{(C7)} 
\begin{equation}
    \lim_{m\rightarrow\infty} \norm{\Delta\etaLiftsim_{(m)}}_{\infty,1}=0
\end{equation}
which implies
\begin{equation}
    \lim_{m\rightarrow\infty}\norm{\Delta\etamodsim_{
(m)}(k)}_\infty=0\,\,\forall k
\end{equation}
$\therefore$ $\forall k$ the sequence $\{\etamodsim_{m}(k)\}$ is a Cauchy sequence, and thus the fixed point $\etamodsim(k)=\lim_{m\rightarrow\infty}\etamodsim_{(m)}(k)$ is unique.
\end{IEEEproof}

\begin{remark}
Neither the preceding presentation nor 
the nonlinear stable inversion prior art
\cite{Zeng2000} 
explicitly
discusses
the intuitive foundation of stable inversion: evolving the stable modes of an inverse system forwards in time from an initial condition and evolving the unstable modes backwards in time from a terminal condition. Unlike for linear time invariant (LTI) systems, this intuition is not put into practice directly for nonlinear systems because 
the similarity transforms that completely decouple the stable and unstable modes of linear systems do not necessarily decouple the stable and unstable modes of nonlinear systems.
However, the same principle underpins this work. This is evidenced by the fact that the intuitive LTI stable inversion is recovered from (\ref{eq:implicitsoln}) when $\fmod$ is LTI, as illustrated briefly below.

For LTI $\fmod$, $\fmodsim$ takes the form
\begin{align}
    \etamodsim(k+1)&=
    \Asim\etamodsim(k) + \Bsim\ypreview(k)
    \\
    \begin{bmatrix} \etamodsim_{s}(k+1) \\ \etamodsim_{u}(k+1) \end{bmatrix}
    &=
    \begin{bmatrix}
    \Asim_s & 0 \\ 0 & \Asim_u 
    \end{bmatrix}\begin{bmatrix}
    \etamodsim_{s}(k) \\ \etamodsim_{u}(k)
    \end{bmatrix} + \begin{bmatrix}
    \Bsim_s \\ \Bsim_u
    \end{bmatrix}\ypreview(k)
\end{align}
Then the implicit solution (\ref{eq:implicitsoln}) becomes the explicit solution
\begin{align}
    \etamodsim(k) &= \sum_{\idx=-\infty}^\infty \phi(k-\idx)\Bsim\ypreview(\idx-1)
    \\
    \begin{bmatrix}
    \etamodsim_s(k) \\ \etamodsim_u(k)  
    \end{bmatrix}
    &= \begin{bmatrix}
    \sum_{\idx=-\infty}^{k}\Asim^{k-\idx}_s\Bsim_s\ypreview(\idx-1)
    \\
    -\sum_{\idx=k+1}^\infty \Asim_u^{k-\idx}\Bsim_u\ypreview(\idx-1)
    \end{bmatrix}
    =
    \begin{bmatrix}
    \Asim_s\etamodsim_s(k-1) + \Bsim_s\ypreview(k-1)
    \\
    \Asim_u^{-1}\etamodsim_u(k+1)-\Asim_u^{-1}\Bsim_u\ypreview(k)
    \end{bmatrix}
\end{align}
which is the forward evolution of the stable modes and backward evolution of the unstable modes where the initial and terminal conditions at $k=\pm\infty$ are zero.

\end{remark}

\subsection{
Initial Picard Iterate 
\texorpdfstring{$\etamodsim_{(0)}$}{} Selection and Implementation
}
This subsection addresses the need to select a new initial Picard iterate $\etamodsim_{(0)}(k)$ in the absence of \ref{(Z3)}.
Also addressed is
 the fact that (\ref{eq:picardinf}) is a purely theoretical, rather than implementable, solution because it contains infinite sums along an infinite timeline.

In the context of ILC, the learned feedforward control action is often intended to be a relatively minor adjustment to the primary action of the feedback controller. 
Thus, choosing $\etamodsim_{(0)}(k)$ to be the feedback-only trajectory, i.e. the zero-feedforward-input trajectory, is akin to warm-starting the fixed-point solving process. This trajectory is given by
\begin{equation}
\begin{aligned}
    \xmod(k+1) &= \fmod\left( \xmod(k), 0, k \right) \qquad \xmod(0)=0_{n_x}
    \\
    \etamodsim_{(0)}(k) &=     V^{-1}
    \begin{bmatrix}
    0_{\mu\times\mu} 
    & 0_{\mu\times n_{\eta}} \\ 0_{n_{\eta}\times \mu} & I_{n_{\eta}\times n_{\eta}}
    \end{bmatrix}\xmod(k)
    \end{aligned}
    \label{eq:picard0}
\end{equation}
for $k\in\{0,\cdots,N-\mu\}$.

An implementable version of (\ref{eq:picardinf}) is given by
\begin{equation}
    \etamodsim_{(m+1)}(k) =
    \sum_{i=1}^{N-\mu+1}
    \phi(k-i)
    \left(\fetasim\left(\etamodsim_{(m)}(i-1),\ypreview(i-1),i-1\right) - \Asim\etamodsim_{(m)}(i-1) \right)
    \label{eq:implementableStabInv}
\end{equation}
for $k\in\{1,...,N-\mu\}$,
fixing the initial condition
$\etamodsim_{(m)}(0)=0_{n_{\eta}}$ 
$\forall m$.

Note that (\ref{eq:implementableStabInv}) is equivalent to 
assuming $\etamodsim_{(m)}(k)=0$, 
$\ypreview(k)=0$,
and $\fetasim\left(0,0,k\right)=0$
for $k\in(-\infty,-1]\cup[N-\mu+1,\infty)$
and
extracting the $k\in[1,N-\mu]$ elements of $\etamodsim_{(m+1)}(k)$ generated by (\ref{eq:picardinf}).
These assumptions correspond to a lack of control action prior to $k=0$ and a reference trajectory that brings the system back to its zero initial condition 
with enough trailing zeros for the system to settle by $k=N-\mu$.
This is typical of repetitive motion processes, but admittedly may preclude some other ILC applications.

Furthermore, 
for the first Picard iteration ($m+1=1$)
these assumptions yield identical (\ref{eq:implementableStabInv})-generated and (\ref{eq:picardinf})-generated $\etamodsim_{(1)}(k)$ on $k\in[0,N-\mu]$.
Because output tracking of 
systems with unstable inverses
typically
 requires preactuation,
for this range of $k$ to contain a practical control input trajectory there must be sufficient leading zeros in the reference starting at $k=0$.
For the following Picard iterates the theoretical and implementable trajectories are unlikely to be equal, but can be made closer the more leading zeros are included in the reference.

Ultimately, applying (\ref{eq:implementableStabInv}) for any number of iterations 
$\mfinal\geq 1$
yields an expression for each time step of 
$\etamodsim_{(\mfinal)}(k)$
whose only variable parameters are the elements of $\yLiftmod$. This is because the recursion calling 
$\etamodsim_{(\mfinal)}(k)$
terminates at the known trajectory $\etamodsim_{(0)}(k)$, and because $\ymod(k)=0$ for $k\in\{0,...,\mu-1\}$ due to the known initial condition $\xmod(0)=0$. The concatenation of these expressions plugged into 
the inverse output function 
(\ref{eq:invout}) yields the lifted inverse system model
\begin{equation}
    \gmod^{-1}(\yLiftmod)=\begin{bmatrix}
    \fmod^{\mu}{\vphantom{\fmod}}^{ -1 }\left(
    V
    \etamodsim_{(\mfinal)}(0),
    \ypreview(0),0\right)
    \\
    \fmod^{\mu}{\vphantom{\fmod}}^{ -1 }\left(
    V
    \etamodsim_{(\mfinal)}(1) ,
    \ypreview(1),1\right)
    \\ \vdots \\
    \fmod^{\mu}{\vphantom{\fmod}}^{ -1 }\left(
    V
    \etamodsim_{(\mfinal)}(N-\mu),
    \ypreview(N-\mu),N-\mu\right)
    \end{bmatrix}
    \label{eq:ginv}
\end{equation}
which enables the 
synthesis of the 
ILILC
learning matrix (\ref{eq:gammanew}).
With this, the complete synthesis 
of ILILC with stable inversion---starting from a model 
in the normal form (\ref{eq:simmuall})---can
be summarized by Procedure \ref{proc:synth}.

\begin{algorithm}
	\caption{ILILC Synthesis with Stable Inversion}
		\label{proc:synth}
\begin{algorithmic}[1]

\State
\label{step:minimalinverse}
Derive the minimal state space representation $\fmod_{\eta}$ and $\fmod^{\mu}{\vphantom{\fmod}}^{-1}$ (from (\ref{eq:invall})) of the conventional inverse of (\ref{eq:simmuall}).

\State
Apply similarity transform $V$ (from (\ref{eq:transform})) to derive the inverse state dynamics representation $\fetasim$ (from (\ref{eq:fetasim})) with decoupled stable and unstable linear parts.

\State
\label{step:etasim}
Use the fixed-point problem solver (\ref{eq:picard0})-(\ref{eq:implementableStabInv}) to derive the inverse system state $\etamodsim_{(\mfinal)}$ as a function of $\yLiftmod$ at each point in time $k\in\{0,\cdots,N-\mu\}$.

\State
\label{step:ginv}
Derive the lifted inverse model $\gmod^{-1}$ 
via (\ref{eq:ginv}).

\State
\label{step:AD}
Use an automatic differentiation tool to 
derive $\jacobian{\gmod^{-1}}{\yLiftmod}$ as a function of $\yLift$, i.e. the learning matrix $L_\tdx$ from (\ref{eq:gammanew}).

\State
\label{step:intertrial}
Compute
$L_\tdx = \jacobian{\gmod^{-1}}{\yLiftmod}(\yLift_\tdx)$ at each trial for the ILC law (\ref{eq:ILCclassic}).

\Comment{
Steps \ref{step:etasim}-\ref{step:ginv} are greatly facilitated by using a computer algebra system. CasADi can provide this functionality in addition to automatic 
differentiation.
}

\end{algorithmic}
\end{algorithm}

\begin{remark}
The computation time required to synthesize ILILC with stable inversion grows with the number of time steps $N$ in the time series, and can become relatively long. However, the overwhelming majority of this computation is performed before the execution of the zeroth trial and need not be repeated. This allows for minimal computation time---i.e. minimal downtime---between trials.

More specifically, Steps \ref{step:minimalinverse}-\ref{step:AD} are all performed before trial zero execution, with Step \ref{step:AD} being the most computationally intensive. These steps yield a function $\jacobian{\gmod^{-1}}{\yLiftmod}(\cdot)$ that arithmetically produces a learning matrix $L_\tdx$ given an output time series.
Step \ref{step:intertrial}---the only step featuring intertrial computation---merely needs to call this function and the simple matrix-vector multiplication of (\ref{eq:ILCclassic}). The fixed-point problem solving and automatic differentiation does not need to be redone.

For reference, 
the validation system's computation times for each step of Procedure \ref{proc:synth} are given in Section \ref{sec:results}, Table \ref{table:computationtime}.
\end{remark}

\section{Validation}
\label{sec:Val}

This section presents validation of the fundamental claim that the original NILC fails for 
models with unstable inverses 
and that the 
newly
proposed
ILILC
framework---when used with stable inversion---succeeds.
Additionally, while the intent of ILC is to account for model error, overly erroneous modeling can cause
violation of \ref{C3}, which 
may cause 
divergence of the ILC law.
Thus this section also probes the performance and robustness of 
ILILC 
with stable inversion over increasing model error in physically motivated simulations.

The 
ILILC law
(\ref{eq:ILCclassic}), (\ref{eq:gammanew}) is applied as a reference shaping tool to a feedback control system (sometimes called ``series ILC''). 
This represents the common scenario of applying a higher level controller to ``closed source'' equipment.
The resultant system (\ref{eq:modelall}) is a nonlinear time-varying system with relative degree $\mu=2$.

Modeling error
is simulated by synthesizing the ILC laws from a nominal ``control model'' of the example system, and applying the resultant control inputs to a set of ``truth models'' featuring random parameter errors and the injection of process and measurement noise.
Finally, to give context to the results for 
ILILC
with stable inversion, identical simulations are run with a benchmark technique that does not require modification for models with unstable inverses: gradient ILC.

\subsection{Benchmark Technique: Gradient ILC}

Gradient ILC is gradient descent applied to the optimization problem
\begin{equation}
    \argmin_{\uLift}\frac{1}{2}\eLift^T\eLift
\end{equation}
which yields the ILC law
\begin{equation}
    \uLift_{\tdx+1} = \uLift_\tdx + \gamma \jacobian{\gmod}{\uLift}\left(\uLift_\tdx\right)^T\eLift_j
    \label{eq:gradILC}
\end{equation}
where $\gamma>0$ is the gradient descent step size. Note that (\ref{eq:gradILC}) is free of the matrix inversion that inhibits the application of NILC  to 
systems with unstable inverses.
Past work on gradient ILC\cite{Owens2009} has been limited to linear systems due in part to the difficulty of synthesizing $\jacobian{\gmod}{\uLift}$ for nonlinear systems. This article extends gradient ILC to nonlinear systems by using the automatic differentiation tool CasADi to synthesize $\jacobian{\gmod}{\uLift}$.

The tuning parameter $\gamma$
influences the performance-robustness trade off of (\ref{eq:gradILC}). 
Reducing $\gamma$ improves the probability that (\ref{eq:gradILC}) will converge for some unknown model error, but may also reduce the rate of convergence. 
For the sake of comparing the convergence rates between gradient ILC and 
ILILC, here we choose $\gamma$ such that the two methods have comparable probabilities of convergence over the 
set
of random model errors tested: $\gamma=1.1$.

\subsection{Example System}
\label{sec:exSystem}
\begin{figure}
    \centering
    \includegraphics[scale=0.3]{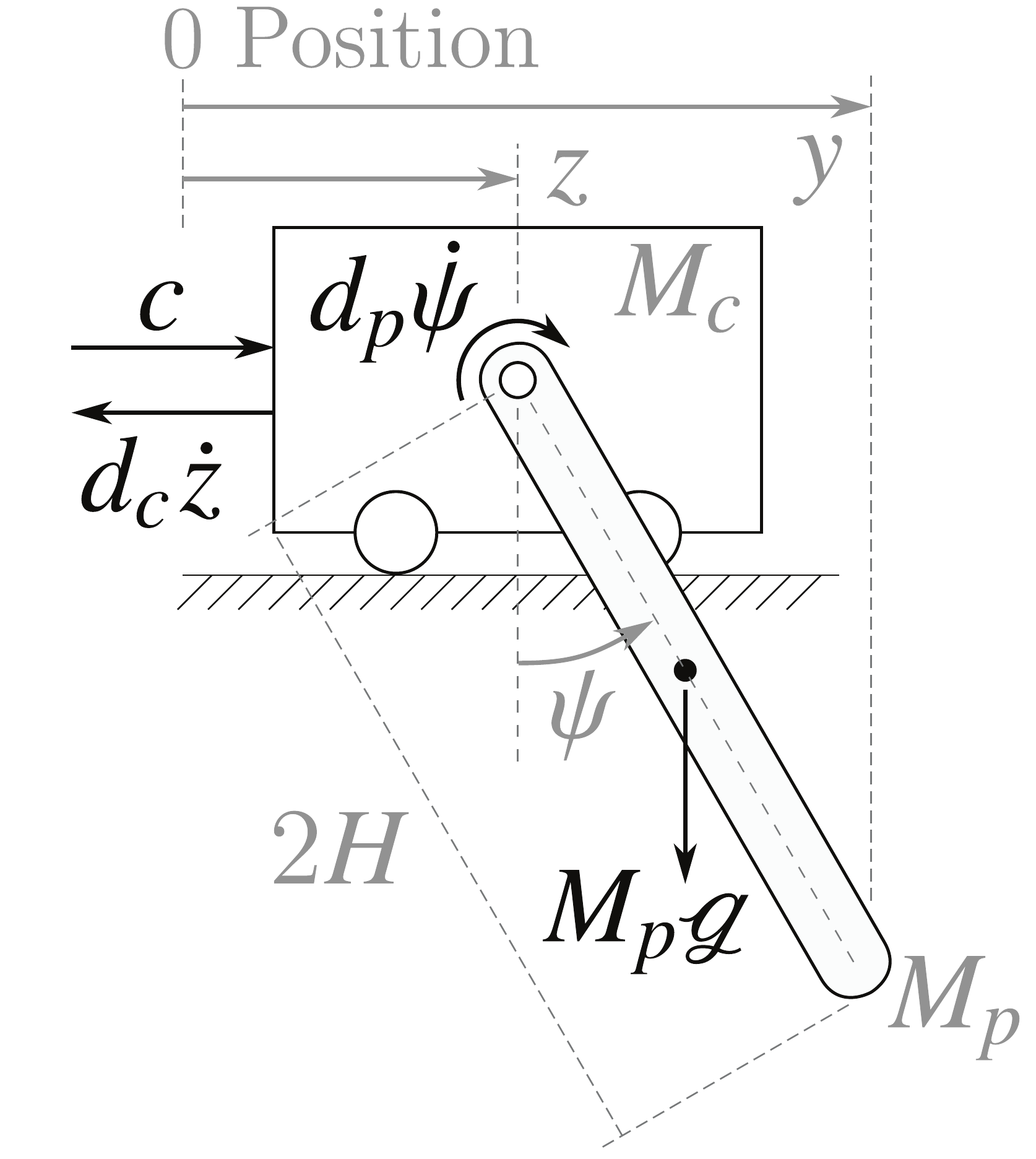}
    \caption{Cart and pendulum system. Dimension, position, and mass annotations are in grey. Force and torque annotations are in black.}
    \label{fig:cartpend}
\end{figure}

Consider the system pictured in Figure \ref{fig:cartpend}, consisting of a pendulum fixed to the mass center of a cart on a rail. 
This subsection presents the first-principles continuous-time equations of motion for this plant, the method for converting these dynamics to the discrete-time normal form (\ref{eq:simmuall}), and the control architecture of the system.

The cart is subjected to an applied force $\force$, and
viscous damping occurs both between the cart and the rail and between the pendulum and the cart.
Equations of motion for this plant 
are given by
\begin{multline}
\label{eq:EOMconttheta}
    \ddot{\angtruth} =
     -3 \left(
     H \Mp \left( \force + \omega_\force \right) \cos (\angtruth ) +
     \dptruth ( \Mc + \Mp ) \dot{\angtruth}+
      H^2 \Mp^2 \sin (\angtruth ) \cos (\angtruth ) \dot{\angtruth}^2+
     \grav H \left(\Mc \Mp +\Mp^2\right) \sin (\angtruth )
    \vphantom{\left(\Mc \Mp +\Mp^2\right)}
    \right. \\ \left.
    \vphantom{
    \left(\Mc \Mp +\Mp^2\right)
     }
     -
  \dctruth H \Mp \cos (\angtruth )\dot{z}
  \right)
  \frac{1}{ H^2 \Mp \left(
  4 (\Mc+\Mp)-3 \Mp \cos ^2(\angtruth )
  \right) }
  \end{multline}%
\begin{multline}
    \label{eq:EOMcontx}
    \ddot{z} =
    \left(
    4 H \left( \force + \omega_\force \right) +
     3 \dptruth  \cos (\angtruth ) \dot{\angtruth}+
     4  H^2 \Mp \sin (\angtruth ) \dot{\angtruth}^2+
     3 \grav H\Mp \sin (\angtruth ) \cos (\angtruth ) 
     \right.\\\left. 
     \vphantom{
     4 H \left( \force + \omega_\force \right) +
     3 \dptruth  \cos (\angtruth ) \dot{\angtruth}+
     }
     - 4 \dctruth  H \dot{z}
     \right)
  \frac{1}{
    H \left(4 (\Mc+\Mp) -3 \Mp \cos ^2(\angtruth )\right)
  }
  \end{multline}
  where
  $\angtruth(k)$ is the pendulum angle, $z(k)$ is the cart's horizontal position, $\grav=\SI{9.8}{\meter\per\second\squared}$ is gravitational acceleration, 
  and the process noise $\omega_\force(k)$ is a random sample from a normal distribution with $0$ mean and standard deviation \SI{3.15e-2}{\newton}. 
  $H$ is the pendulum half-length, $\Mc$ and $\Mp$ are the cart and pendulum masses, and $\dctruth$ and $\dptruth$ are the cart-rail and pendulum-cart damping coefficients, respectively.
  The time argument of $\omega_c$, $\angtruth$, $z$ and their derivatives has been dropped for compactness.

The output to be tracked is the pendulum tip's horizontal position, $y$.
Obtaining a discrete-time state space model of this system in 
the
normal form
(\ref{eq:simmuall})
requires first a change of coordinates
such that the desired output is a state,
and then discretization. The change of coordinates is
\begin{equation}
    \angtruth = \arcsin\left(\frac{y-z}{2H}\right)
    \label{eq:ChangeOfCoord}
\end{equation}
with
associated derivative substitutions
\begin{align}
    \dot{\angtruth} &= \frac{\dot{y}-\dot{z}}{2H\sqrt{1-\frac{\left(y-z\right)^2}{4H^2}}}
    \\
    \ddot{\angtruth} &=
    \frac{
    \sec\left(\angtruth\right)\left(
    \ddot{y}-\ddot{z} + 2H\sin\left( \angtruth \right)\dot{\angtruth}^2
    \right)
    }{
    2H
    }
    \label{eq:ddang}
\end{align}
Then the equations of motion are solved for in terms of the new coordinates.
In the present case
(\ref{eq:EOMconttheta})-(\ref{eq:ddang}) can be solved for $\ddot{y}(k)$ and $\ddot{z}(k)$ as functions of $y(k)$, $z(k)$, $\dot{y}(k)$, and $\dot{z}(k)$.
Next, forward Euler discretization 
is applied recursively to the equations of motion to reformulate the state dynamics in terms of discrete time increments rather than derivatives, as is required by the normal form. The innermost layer of the recursion is the first derivatives
\begin{equation}
\dot{y}(k)=\frac{y(k+1)-y(k)}{T_s}
\qquad 
\dot{z}(k)=\frac{z(k+1)-z(k)}{T_s}
\end{equation}
where 
the sample period 
$T_s=\SI{0.016}{\second}$ in this case.
These can be plugged into 
$\ddot{y}(k)$ and $\ddot{z}(k)$ to eliminate their dependence on derivatives.
The next---and in this case final---layer is
the forward Euler discretization of the second derivatives.
The outermost layer 
can be rearranged to yield the discrete-time equations of motion 
\begin{equation}
\label{eq:EOMdisc}
\begin{aligned}
    y(k+2)&=\ddot{y}(k)T_s^2+2y(k+1)-y(k)
    \\
    z(k+2)&=\ddot{z}(k)T_s^2+2z(k+1)-z(k)
    ,
\end{aligned}
\end{equation}
which are directly used to define the state dynamics $f$ in terms of the state vector
$x(k)=[y(k),\,y(k+1),\,z(k),\,z(k+1)]^T$.
The explicit expressions of (\ref{eq:EOMdisc}) 
are too long to print here, but can be easily obtained in Mathematica, MATLAB symbolic toolbox, etc. via the algebra 
described 
in (\ref{eq:ChangeOfCoord})-(\ref{eq:EOMdisc}).

\begin{figure}
    \centering
    \includegraphics[scale=0.4]{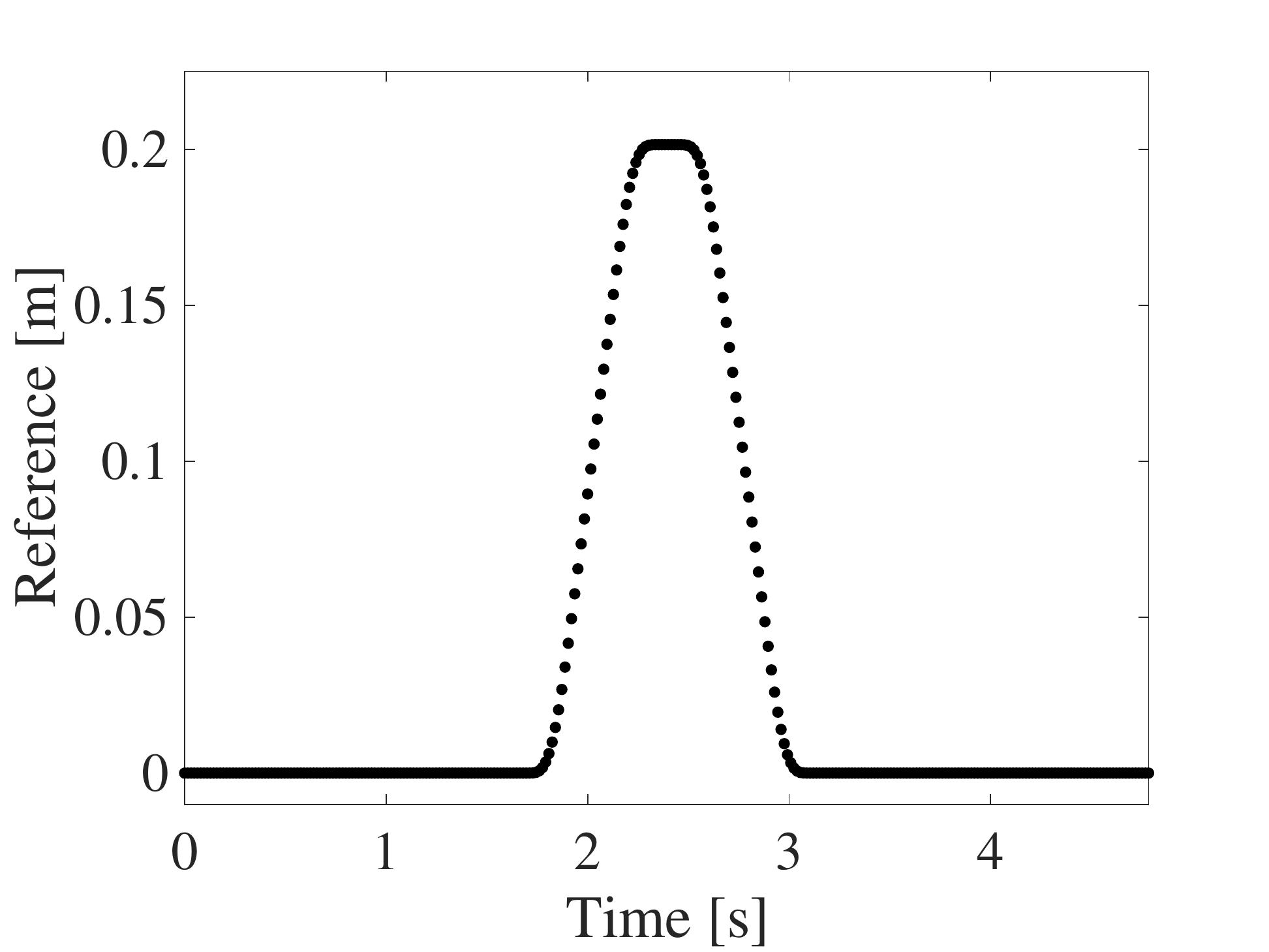}
    \caption{Reference}
    \label{fig:ref}
\end{figure}

The output must track the reference $r(k)$ given in Figure \ref{fig:ref}. To accomplish this the plant is equipped with a full-state feedback controller modeled as
\begin{align}
    \force(k)&= \kappa_0r^*(k) - 
    \begin{bmatrix}
    \kappa_1 & \kappa_2 & \kappa_3 &\kappa_4
    \end{bmatrix}x(k)
    \\
    \rPerceived(k)&= r(k)+u(k)
\end{align}
Here, $\rPerceived(k)$ is the effective reference and $u(k)$ is the control input generated by the ILC law. In other words, the ILC law adjusts the reference delivered to the feedback controller to eliminate the error transients inherent to feedback control.
Finally, the error signal input to the ILC law is subject to measurement noise $\omega_y(k)$
\begin{equation}
    e(k)=r(k)-y(k)-\omega_y(k)
\end{equation}
where the noise's distribution has 0 mean and standard deviation \SI{5e-5}{\meter}.

\begin{figure}
    \centering
    \includegraphics[scale=0.95]{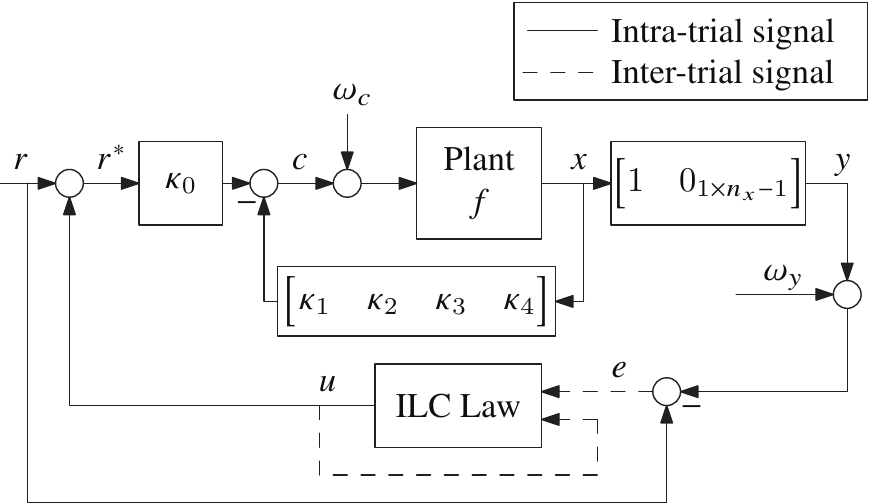}
    \caption{
    System Block Diagram. The control law outputting $u$ is synthesized from the control models defined by the behatted parameters of Table \ref{table:cartpend} and by $\unoise(k)=\ynoise(k)=0$. The plant and controller gain blocks are defined with the truth model parameters generated according to Section \ref{sec:method}.
    Inter-trial signals from trial $\tdx$ are stored and used to compute the input for trial $\tdx+1$.
    }
    \label{fig:block}
\end{figure}

The ILC law itself is synthesized from a 
control model
that is identical in structure to the 
truth model
presented above, but has $\hat{\omega}_\force=\hat{\omega}_y=0$ and uses the model parameters tabulated in Table \ref{table:cartpend}.
Stable inversion for the synthesis of 
learning matrix 
(\ref{eq:gammanew}) is performed with a single Picard iteration, i.e. $\mfinal=1$ in (\ref{eq:ginv}).
To simulate model error, the hatless truth model parameters differ from the behatted control model parameters in a manner detailed in Section \ref{sec:method}.
This ultimately results in the system block diagram given in Figure \ref{fig:block}.

\begin{table}
\centering
\caption{
Cart-Pendulum Control Model Parameters
}
\label{table:cartpend}
\renewcommand{\arraystretch}{1.25}
\begin{tabular}{|l|c|c|}
\hline
Parameter & Symbol & Value 
\\
\hhline{|=|=|=|}
Cart Mass & $\Mcmod$ & \SI{0.5}{\kilo\gram}
\\
\hline
Pendulum Mass & $\Mpmod$ & \SI{0.25}{\kilo\gram}
\\
\hline
Pendulum Half-Length & $\hat{H}$ & \SI{0.225}{\meter}
\\
\hline
Cart-Rail Damping Coefficient & $\dcmod$ & \SI{10}{\kilo\gram\per\second}
\\
\hline
Pendulum-Cart Damping Coefficient & $\dpmod$ & 
\SI{0.01}{\kilo\gram\meter\squared\per\second}
\\
\hline
Full State Feedback Gain 0 & $\kmod_0$ & \SI{630}{}
\\
\hline
Full State Feedback Gain 1 & $\kmod_1$ & \SI{-5900}{}
\\
\hline
Full State Feedback Gain 2 & $\kmod_2$ & \SI{5900}{}
\\
\hline
Full State Feedback Gain 3 & $\kmod_3$ & \SI{-3700}{}
\\
\hline
Full State Feedback Gain 4 & $\kmod_4$ & \SI{4300}{}
\\
\hline
\end{tabular}
\end{table}

\subsection{Simulation and Analysis Methods}
\label{sec:method}
Let $\parammod\in\real^{10}$ be a vector of the control model parameters in Table \ref{table:cartpend}. Then a truth model can be specified by the vector $\theta$, generated via
\begin{equation}
    \param = \left( 1_{10\times 1} e_{\param}^T\odot I + I \right)\parammod
    \label{eq:truthModlGen}
\end{equation}
where $\odot$ is the Hadamard product and $e_{\param}\in\real^{10}$ is a random sample of a uniform distribution. Under (\ref{eq:truthModlGen}), each element of $e_{\param}$ is the relative error between the corresponding elements of $\param$ and $\parammod$. Thus, $\norm{e_{\param}}_2$ provides a scalar metric for the model error between the control model and a given truth model. The range $\norm{e_{\param}}_2\in[0,0.1]$ is divided into 20 bins of equal width, and 50 truth models are generated for each bin. Both ILC schemes are applied to each truth model with 50 trials, 
and $u_0(k)=0$ $\forall k$.
A full set of 50 trials of one of the ILC laws applied to a single truth model is referred to as a ``simulation.''
The results of these simulations are used to characterize the probability of convergence and rate of convergence of each ILC law.

For each iteration of a simulation, the normalized root mean square error (NRMSE) is given by
\begin{equation}
    \text{NRMSE}_\tdx \equiv \frac{\RMS\left(\eLift_\tdx\right)}{\norm{\rLift}_\infty}
\end{equation}
A simulation is deemed convergent if there exists $\tdx^*$ such that $\text{NRMSE}_\tdx$ is less than some tolerance for all $\tdx\geq \tdx^*$. This work uses a tolerance of \SI{5e-4}{}, which is close to the NRMSE floor created by noise.

Let $\tdx^{\beta,\tau,\lambda}$ be the minimum $\tdx^*$ for truth model $\tau\in[1,50]$ in bin $\beta\in[1,20]$ under ILC law $\lambda\in\{\text{
ILILC
},\,\text{gradient ILC}\}$, and
let $\convset$ be the set of all $(\beta,\tau)$ for which both 
ILILC 
and gradient ILC converge. Then
the mean transient convergence rate 
\begin{equation}
    \convRate_{\lambda}=\mean_{\convset,\tdx\in[1,\tdx^{\beta,\tau,\lambda}]}
    \left( \frac{\text{NRMSE}^{\beta,\tau,\lambda}_{\tdx}}{\text{NRMSE}^{\beta,\tau,\lambda}_{\tdx-1}} \right)
    \label{eq:convRate}
\end{equation}
offers a numerical performance metric.
Note that Avrachenkov\cite{Avrachenkov1998} gives a theoretical convergence analysis for the ILC structure (\ref{eq:ILCclassic}) in general (covering 
NILC,
ILILC,
and gradient ILC). This analysis can be used to lower bound performance (i.e. upper bound convergence rate) via multiple parameters computed from the learning matrix $L_\tdx$ and the true dynamics $\gtrue$. The mean transient convergence rate (\ref{eq:convRate}) may thus serve as a specific, measurable counterpart to any theoretical worst-case-scenario analyses performed via the formulas in the work of Avrachenkov\cite{Avrachenkov1998}.

Finally, to verify the fundamental necessity and efficacy of 
ILILC  for systems with unstable inverses, 
2 trials of
traditional stable-inversion-free NILC  (\ref{eq:ILCclassic}), (\ref{eq:gammaprior})
are applied to each truth model.

All computations are performed on a desktop computer with a \SI{4}{\giga\hertz} CPU and \SI{16}{\giga\byte} of RAM.

\subsection{Results and Discussion}
\label{sec:results}
The condition number of $\jacobian{\gmod}{\uLift}\left(\uLift_0\right)$ is \SI{1e17}{}.
Attempted inversion of this matrix in MATLAB 
yields an inverse matrix with average nonzero element magnitude of \SI{4e13}{} and max element magnitude of \SI{3e16}{}. Consequently, $\uLift_1$ generated by (\ref{eq:ILCclassic}), (\ref{eq:gammaprior}) has an average element magnitude of \SI{2e10}{\meter} and a max element magnitude of \SI{8e11}{\meter}, which is so large that $\yLift_1$ and $\jacobian{\gmod}{\uLift}\left(\uLift_1\right)$ contain \texttt{NaN} elements for all simulations.
Conversely, while some simulations using 
ILILC , i.e. 
(\ref{eq:ILCclassic}), (\ref{eq:gammanew}), diverge due to excessive model error, the majority converge. 
Additionally, the computation times given in Table \ref{table:computationtime} show that Procedure \ref{proc:synth} successfully front-loads almost all of the 
required computation;
intertrial computation time is almost always less than \SI{150}{\milli\second}. 
Together, these results validate
the fundamental claim that the direct application of Newton's method in NILC  is insufficient for 
systems with unstable inverses, 
and that 
the combination of ILILC and stable inversion
fills
this gap.

\begin{table}
    \centering
    \caption{
    Computation Times for the Steps of Procedure \ref{proc:synth}
    }
    \label{table:computationtime}
    \setstretch{1.2}
\begin{tabular}{|l|c|c|rl|}
\hline
When Performed                                                                                 & Step                                    & Operation                                                                                                                                                              & \multicolumn{2}{c|}{Time [\SI{}{\second}]}        \\ \hhline{|=|=|=|==|}
\multirow{2}{*}{\begin{tabular}[c]{@{}l@{}}Once, before \\execution of trial 0 \end{tabular}}  & 1-2                                     & Derive the minimal, similarity transformed inverse state space system                                                                                                  & \multicolumn{2}{c|}{\SI{6}{}}                     \\ \cline{2-5}  
                                                                                               & 3                                       & \begin{tabular}[c]{@{}c@{}}Fixed-point-problem-solving-based stable inversion\\ to derive the inverse state trajectory as a function of $\yLiftmod$\end{tabular}       & \multicolumn{2}{c|}{\SI{95}{}}                    \\ \cline{2-5}
                                                                                               & 4                                       & Conversion to lifted input-output model $\gmod(\cdot)$                                                                                                                 & \multicolumn{2}{c|}{\SI{0.3}{}}                   \\ \cline{2-5}
                                                                                               & 5                                       & Automatic differentiation to produce $\jacobian{\gmod^{-1}}{\yLiftmod}(\cdot)$                                                                                         & \multicolumn{2}{c|}{\SI{573}{}}                   \\ \hline
\multirow{2}{*}{\begin{tabular}[c]{@{}l@{}}Between trials\\ 49,000 samples\end{tabular}}       & \multicolumn{1}{c|}{\multirow{2}{*}{6}} & \multicolumn{1}{l|}{\multirow{2}{*}{Update of feedforward input trajectory via $L_\tdx = \jacobian{\gmod^{-1}}{\yLiftmod}(\yLift_\tdx)$ and (\ref{eq:ILCclassic})}}    & Mean:           & 0.138                           \\
                                                                                               & \multicolumn{1}{l|}{}                   & \multicolumn{1}{l|}{}                                                                                                                                                  & Std:            & 0.006                           \\ \hline
\end{tabular}
\end{table}

\begin{figure}
    \centering
    \includegraphics[scale=0.9]{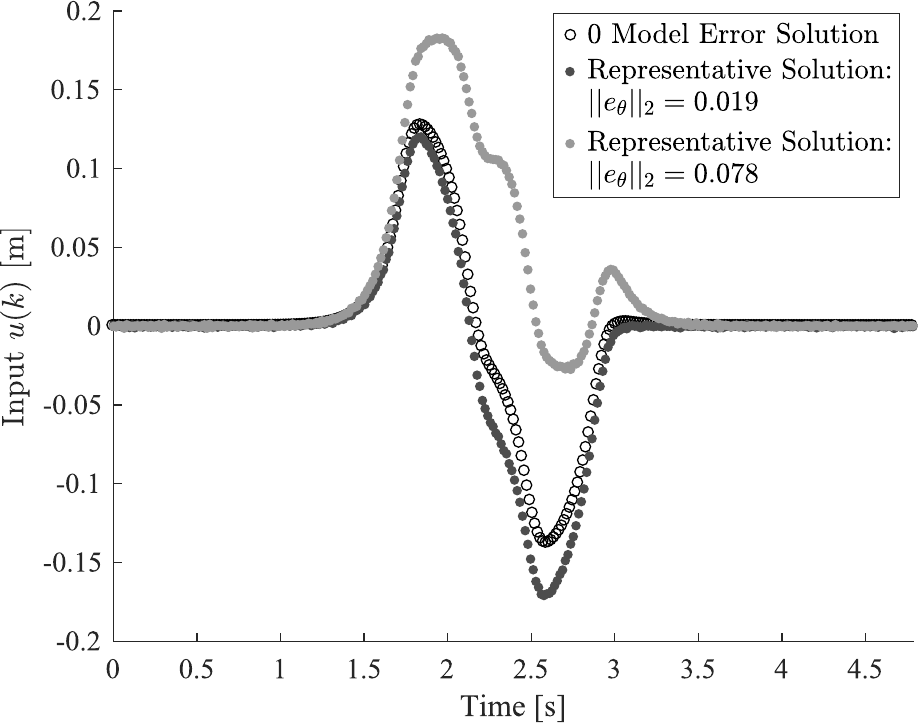}
    \caption{
    Representative input solution trajectories from low- and high-model-error 
    ILILC  
    simulations compared with the solution to the zero-model-error problem.
    The zero-model-error solution is the input trajectory that would be chosen for feedforward control in the absence of learning, and differs notably from both minimum-error trajectories found by 
    ILILC   
    with stable inversion.
    }
    \label{fig:solution}
\end{figure}

To accompany the quantitative metric $\norm{e_\theta}_2$,
Figure \ref{fig:solution} 
offers a qualitative sense of the degree of model 
error
in this study by
comparing two representative 
ILILC  
solution trajectories $u_{50}(k)$ with the solution to the $\norm{e_\theta}_2=0$, $\omega_\force(k)=\omega_y(k)=0$ scenario.
The lower-model-error representative solution is from within the range of $\norm{e_\theta}_2$ for which all simulations converged, while the higher-model-error solution comes from a bin in which some simulations diverged.
A more detailed analysis of the boundaries in $\theta$-space determining convergence or divergence of a simulation is beyond the scope of this work.
However, the given trajectories illustrate that even in the conservative subspace defined by the 100\% convergent bins learning bridges a visible performance gap, and that beyond this subspace there are far greater performance gains to be had.

\begin{figure}
    \centering
    \includegraphics[scale=0.925,trim={0.14in 0in 0.3in 0.1in},clip]{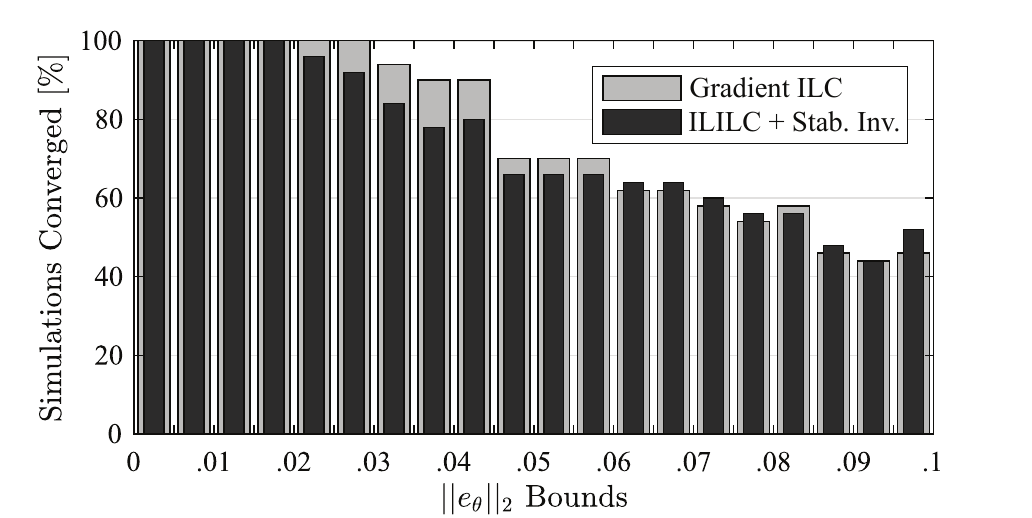}
    \hspace{2mm}
    \includegraphics[scale=0.925,trim={0.14in 0in 0.3in 0.1in},clip]{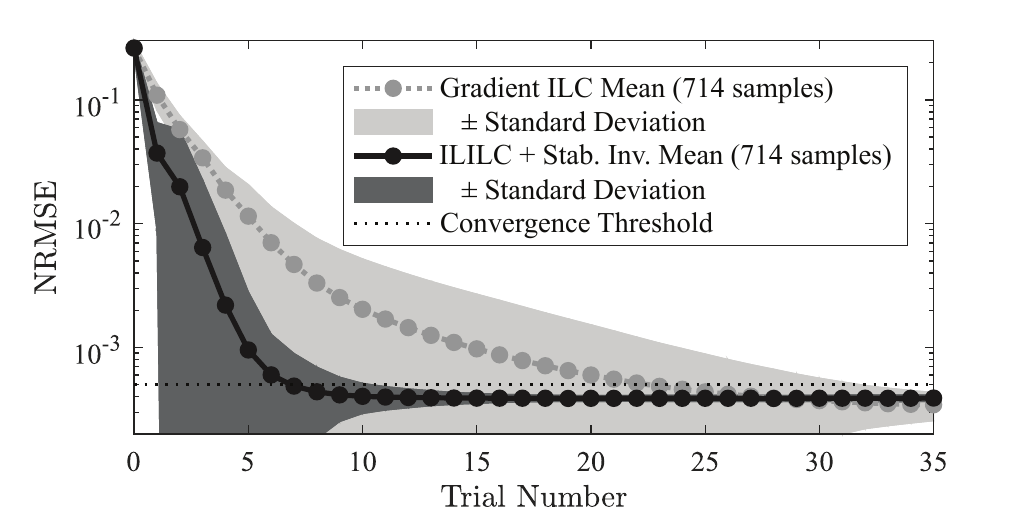}
    \caption{
    \emph{Left:} 
    Histogram giving the percentage of simulations converged 
    in each bin of the model error metric $\norm{e_\theta}_2$.
    \emph{Right:}
    Mean value of NRMSE for each ILC trial over all simulations that are convergent for both gradient ILC and 
    ILILC 
    with stable inversion.
    This illustrates that for comparable robustness to model error, 
    ILILC  
    converges substantially faster than gradient ILC.
    }
    \label{fig:data}
\end{figure}

Finally, a statistical
comparison
of the performance and robustness of 
ILILC  
with stable inversion and gradient ILC is given in Figure \ref{fig:data}.
The tuning of gradient ILC indeed yields comparable robustness to 
ILILC,
with 
ILILC  
97\%
as likely to converge as gradient ILC over all simulations.
The convergence rates of the two ILC schemes, however, differ substantially, with gradient ILC taking over 3 times as many trials as 
ILILC  
to converge on average.
The mean transient convergence rate values tabulated in Table \ref{table:convRate} give a more portable quantification of 
ILILC's 
advantage,
having a convergence rate 
nearly half
that of gradient ILC's.

This analysis confirms that 
ILILC  
with stable inversion is an important addition to the engineer's toolbox because it enables ILC synthesis from nonlinear non-minimum phase models and delivers the fast convergence characteristic of algorithms based on Newton's method.

\begin{table}
\caption{Transient Convergence Rates for 
ILILC 
and Gradient ILC}
\centering
\renewcommand{\arraystretch}{1.25}
\begin{tabular}{|l|c|c|}
\hline
ILC Law           & \multicolumn{1}{l|}{Mean} & \multicolumn{1}{l|}{Standard Deviation} \\ \hhline{|=|=|=|}
Gradient ILC      & 0.76                      & 0.17                                    \\ \hline
ILILC 
+ Stab. Inv. & 0.41                      & 0.27                                    \\ \hline
\end{tabular}
\label{table:convRate}
\end{table}

\section{Conclusion}
\label{sec:conc}
This work introduces and validates 
a new 
ILC synthesis scheme applicable to nonlinear time-varying 
systems with unstable inverses and 
relative degree greater than 1.
 This is done with the support of nonlinear stable inversion, which is 
 advanced from the prior art via proof of convergence for an expanded class of systems and methods for improved practical implementation.
In all, this results in a new, broadly implementable ILC scheme 
displaying
a competitive convergence speed under benchmark testing.

Future work may focus on
further broadening the applicability of 
ILILC  
by relaxing reference and initial condition repetitiveness assumptions, and on
the extension of 
ILILC  
with a potentially adaptive tuning parameter or other means to enable the exchange of some speed for robustness when called for.
Levenberg-Marquardt-Fletcher algorithms may offer one source of inspiration for such work.

%\backmatter

\section*{Acknowledgments}
This work is supported by the U.S. National Science Foundation (CAREER Award \#1351469), the U.S. Department of Commerce, National Institute of Standards and Technology (Award 70NANB20H137), and the Netherlands Organization for Scientific Research (research programme VIDI project 15698)

% \subsection*{Author contributions}

% This is an author contribution text. This is an author contribution text. This is an author contribution text. This is an author contribution text. This is an author contribution text. 

% \subsection*{Financial disclosure}

% None reported.

\subsection*{Conflict of interest}

The authors declare no potential conflict of interests.

% \section*{Supporting information}

\appendix
\section{: Failure of other ILC schemes for systems with unstable inverses}
\label{app:priorart}

This appendix demonstrates that the sufficient conditions for convergence proposed by 
past works
\cite{Jang1994,Saab1995,Wang1998,Sun2003,Zhang2019}
on ILC for discrete-time nonlinear systems are in actuality not sufficient for
at least
some cases of systems having unstable inverses.
This is done by running model-error-free ILC simulations that are guaranteed to converge by the past works, and observing them to diverge
instead.

\setcitestyle{citesep={, }}% for textual, non-superscript references
Each of references \citenum{Jang1994,Saab1995,Wang1998,Sun2003,Zhang2019} proposes sufficient conditions for the 
convergence 
$\lim_{\tdx\rightarrow\infty}\eLift_\tdx=0_{N-\mu+1}$
of a particular ILC scheme applied to a particular class of nonlinear dynamics.
All of these classes of nonlinear dynamics are supersets of the SISO LTI dynamics
\begin{IEEEeqnarray}{RL}
\eqlabel{eq:LTI}
\IEEEyesnumber
\IEEEyessubnumber*
    x_\tdx(k+1) &= Ax_\tdx(k) + Bu_\tdx(k)
    \\
    y_\tdx(k) &= Cx_\tdx(k)
\end{IEEEeqnarray}
with relative degree $\mu=1$, i.e. $CB\neq 0$. Additionally, assume (\ref{eq:LTI}) is stable and $x_\tdx(0)$ is such that $y_\tdx(0)=r_\tdx(0)$ $\forall\tdx$.
Given a system of this 
structure, the ILC schemes and 
convergence conditions of 
the
past works
reduce to the following.

From reference \citenum{Jang1994} the learning law is
\begin{equation}
    u_{\tdx+1}(k) = u_\tdx(k) + L_\tdx(k)\left(\gamma_1e_\tdx(k+1) + \gamma_0e_\tdx(k) \right)
    \label{eq:JangLearn}
\end{equation}
where $L\in\real$ is a potentially time-varying and trial-varying part of the learning gain and $\gamma_1$, $\gamma_0\in\real$ are trial-invariant, time-invariant learning gains with $\gamma_1\neq 0$. The learning laws of references \citenum{Saab1995,Wang1998,Sun2003,Zhang2019} are special cases of (\ref{eq:JangLearn}): reference \citenum{Saab1995} sets $\gamma_1=1$, $\gamma_0=-1$, reference \citenum{Wang1998} sets $L$ to be trail-invariant, $\gamma_1=1$, $\gamma_0=0$, reference \citenum{Sun2003} sets $\gamma_1=1$ and leaves $\gamma_0$ free, and reference \citenum{Zhang2019} sets $L$ to be trial-invariant and time-invariant, $\gamma_1=1$, $\gamma_0=0$.

Each 
work presents a different variation of convergence analysis, but all propose 
a
sufficient condition for the convergence $\lim_{\tdx\rightarrow\infty}\eLift_\tdx=0_{N-\mu+1}$ under their ILC scheme. 
References \citenum{Jang1994,Wang1998,Sun2003} use
\begin{enumerate}[label=(C\arabic*),labelindent=0pt,labelwidth=\widthof{\ref{CJang}},leftmargin=!]
\resume{conditions02}
    \item 
    $
    |
    1-L_\tdx(k)\gamma_1CB
    |
    <1$ $\forall \,k,\,\tdx$ .
    \label{CJang}
\suspend{conditions03}
\end{enumerate}
Reference \citenum{Zhang2019} uses the stricter condition
\begin{enumerate}[label=(C\arabic*),labelindent=0pt,labelwidth=\widthof{\ref{CZhang}},leftmargin=!]
\resume{conditions03}
    \item 
    $0<L_\tdx(k)CB<1$ $\forall \,k,\,\tdx$ .
    \label{CZhang}
\suspend{conditions04}
\end{enumerate}
Finally, reference \citenum{Saab1995} uses the combination of \ref{CJang} and
\begin{enumerate}[label=(C\arabic*),labelindent=0pt,labelwidth=\widthof{\ref{CZhang}},leftmargin=!]
\resume{conditions04}
\item
$\norm{A}>1$
\label{CSaab}
\suspend{conditions05}
\end{enumerate}
where any consistent norm may be chosen for $\norm{\cdot}$.

Consider the example system and learning gain
\begin{equation}
    \begin{aligned}
    A &= \begin{bmatrix}
    -0.3 & -0.79 & 0.53
    \\
     0   &  0.5  & 1
     \\
     0   & -0.36 & 0.5
    \end{bmatrix}
    \qquad
    &B &= \begin{bmatrix}
    0 \\ 0 \\ 1.34
    \end{bmatrix}
    \\
    C &= \begin{bmatrix}
    0.7 & 1.1 & -0.74
    \end{bmatrix}
    \qquad 
    &x_\tdx(0)&=0 \,\,\forall\tdx
    \end{aligned}
    \label{eq:LTIex}
\end{equation}
\begin{equation}
    L_\tdx(k)=0.5(CB)^{-1} \quad \forall \, k,\,\tdx
    \label{eq:LTILex}
\end{equation}
with the reference given in Figure \ref{fig:ref} and the zeroth control input $u_0(k)=0$ $\forall k$.
This system has an unstable inverse.

\begin{figure}
    \centering
    \includegraphics[scale=0.69,trim={0.14in 0in 0.3in 0.1in},clip]{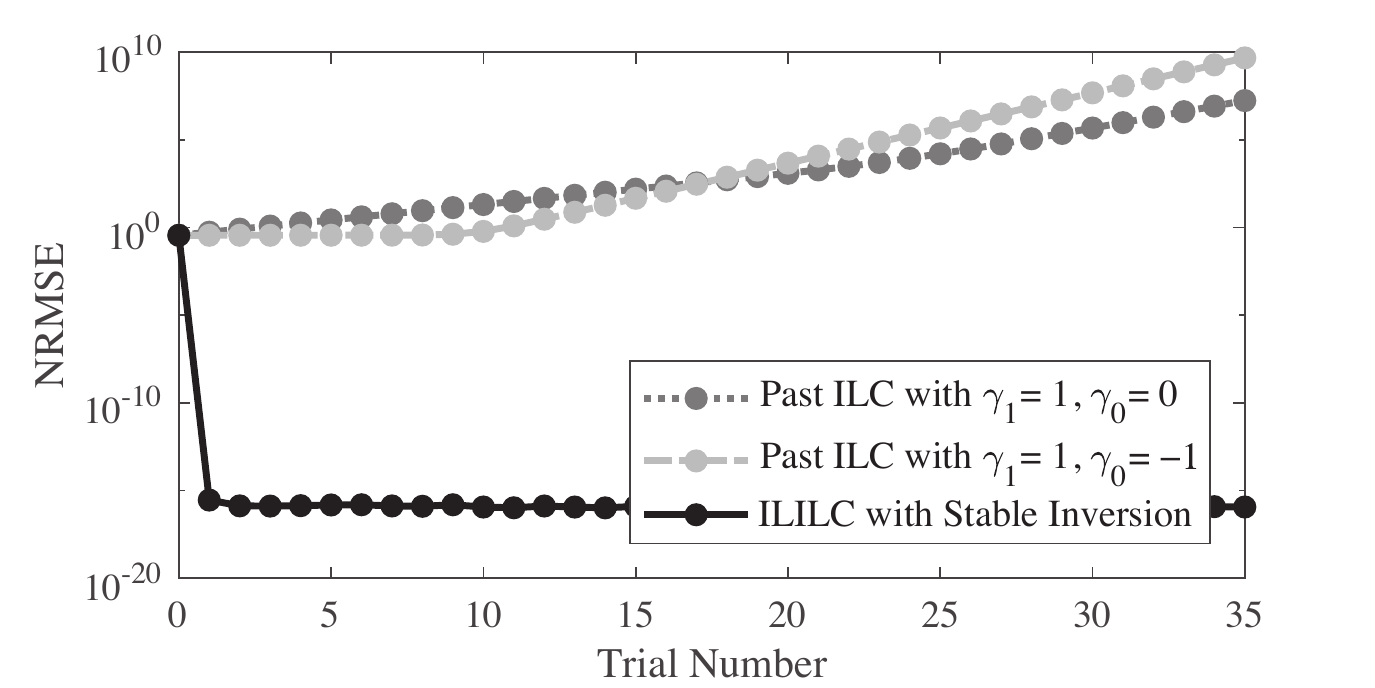}
    \caption{
    NRMSE versus trial number of past works' ILC schemes (\ref{eq:JangLearn}) applied with learning gain (\ref{eq:LTILex})  
    to the system (\ref{eq:LTIex}).
    These NRMSEs monotonically increase, confirming the inability of the past work on ILC with discrete-time nonlinear systems to account for unstable inverses.
    The NRMSE trajectory yielded by the stable-inversion-supported ILILC  scheme proposed by this article is also displayed.
    The convergence of this ILC scheme when applied to (\ref{eq:LTIex}) reiterates its ability to control such non-minimum  phase systems.
    }
    \label{fig:neglect}
\end{figure}

The plant (\ref{eq:LTIex}) satisfies \ref{CSaab}, and with (\ref{eq:LTILex}) it satisfies \ref{CJang} and \ref{CZhang} for $\gamma_1=1$. Thus, according to references \citenum{Jang1994,Saab1995,Wang1998,Sun2003,Zhang2019} the ILC scheme (\ref{eq:JangLearn}) is guaranteed to 
yield tracking error convergence
in a model-error-free simulation.
However, Figure \ref{fig:neglect} shows that 
the tracking error
diverges
under (\ref{eq:JangLearn}),
meaning that satisfaction of \ref{CJang}-\ref{CSaab} is not actually sufficient for the convergence of all systems (\ref{eq:LTI}) under the learning law (\ref{eq:JangLearn}) in practice.
This illustrates that the failure to account for phenomena arising from inverse instability is not unique to NILC, but rather pervades the literature on ILC with discrete-time nonlinear systems.

In light of the counterexample given by (\ref{eq:LTIex}) to the sufficiency of \ref{CJang}-\ref{CSaab} for the convergence of the ILC schemes in references \citenum{Jang1994,Saab1995,Wang1998,Sun2003,Zhang2019}, it is desirable to formalize an additional condition that precludes systems such as (\ref{eq:LTIex}) from consideration for the application of these ILC schemes.
Such a condition is given by:
\begin{enumerate}[label=(C\arabic*),labelindent=0pt,labelwidth=\widthof{\ref{CZhang}},leftmargin=!]
\resume{conditions05}
    \item 
    \label{CLTIInvStab}
    equation (\ref{eq:fetasim}) must be asymptotically stable about its solution.
\end{enumerate}
For SISO LTI systems with relative degree $\mu\geq 1$ (i.e. systems of class (\ref{eq:LTI})), \ref{CLTIInvStab} is equivalent to 
\begin{equation}
\SpecRad\left( A-B\left(CA^{\mu-1}B\right)^{-1}CA^\mu \right)<1
\end{equation}
where $A-B\left(CA^{\mu-1}B\right)^{-1}CA^\mu$ is the state matrix of the inverse system.
(\ref{eq:LTIex}) violates this condition, but many systems satisfy it, including all damped harmonic oscillators discretized via the forward Euler method.
While sufficient, note that \ref{CJang}-\ref{CLTIInvStab} might not be necessary conditions. Analysis of necessary conditions for error convergence under past works' ILC schemes is beyond the scope of this work. As shown in Figure \ref{fig:neglect}, the ILC scheme proposed by the present article is capable of solving the problem presented by the given counterexample---(\ref{eq:LTIex})---to past works' ILC schemes.

\setcitestyle{citesep={,}}% reset to default citation style

% References
\bibliography{spiegelBib}%

% \clearpage

% \section*{Author Biography}

% \begin{biography}{\includegraphics[width=66pt,height=86pt,draft]{empty}}{\textbf{Author Name.} This is sample author biography text this is sample author biography text this is sample author biography text this is sample author biography text this is sample author biography text this is sample author biography text this is sample author biography text this is sample author biography text this is sample author biography text this is sample author biography text this is sample author biography text this is sample author biography text this is sample author biography text this is sample author biography text this is sample author biography text this is sample author biography text this is sample author biography text this is sample author biography text this is sample author biography text this is sample author biography text this is sample author biography text.}
% \end{biography}

\end{document}